\documentclass[twocolumn]{aastex62}

\usepackage{placeins}

\newcommand*{\Cratio}{^{12}\mathrm{C}/^{13}\mathrm{C}}

\begin{document}
\title{Light Element Abundances and Multiple Populations in M53}

\author{Jeffrey M. Gerber}
\affiliation{Max Planck Institute for Astronomy, 17 Königstuhl, Heidelberg, 69117, Germany}
\affiliation{Department of Astronomy, Indiana University Bloomington, Swain West 319, 727 East 3rd Street, Bloomington, IN 47405-7105, USA}

\author{Eileen D. Friel}
\affiliation{Department of Astronomy, Indiana University Bloomington, Swain West 319, 727 East 3rd Street, Bloomington, IN 47405-7105, USA}
\author{Enrico Vesperini}
\affiliation{Department of Astronomy, Indiana University Bloomington, Swain West 319, 727 East 3rd Street, Bloomington, IN 47405-7105, USA}

\begin{abstract}
We present results from a study of 94 red giant stars in the globular cluster M53. We use low-resolution spectra to measure the strength of CN and CH features at $\sim$3800 and 4300 \AA, respectively. The strengths of these features are used to classify stars into a CN-enhanced and CN-normal population and to measure C and N abundances in all 94 stars. We find the red giant branch stars to be evenly split between the two populations identified, and observe the presence of CN-enhanced stars on the asymptotic giant branch. In addition, we identify 5 CH star candidates based on the strength of their CN and CH band features, and the presence of a P-branch in their CH band. We compare our identification of multiple populations to those based on the Na-O anti-correlation and pseudo-color indices in HST UV photometry, and find general agreement between all three methods. Our large sample size also allows us to study the radial distribution of each population, and we find that the CN-enhanced population is more centrally concentrated. We use our C and N measurements to compare the evolutionary changes in these elements as a function of magnitude between the two populations, and show that both populations experience similar evolutionary changes to the surface abundances of C and N. Finally, we calculate C+N+O abundances for each population and compare them to similar measurements made in M10; we find that in both clusters CN-enhanced stars have a slightly enhanced C+N+O ($\Delta$(C+N+O) $\sim$ 0.2 dex).
\end{abstract}


\section{Introduction}
As some of the oldest objects in the Milky Way, globular clusters (GCs) are important to better our understanding of topics such as Galactic formation and evolution, Galactic dynamical history, and low-mass stellar evolution. Numerous studies have revealed a picture of GCs much more complex than that according to which these systems would be composed of chemically homogeneous stars all formed from the same material. 
Early spectroscopic studies of red-giant branch (RGB) stars showed the presence in a few GCs of inhomogeneities in light element abundances such as carbon (C), nitrogen (N), oxygen (O), sodium (Na), aluminum (Al), and magnesium (Mg). After decades of study it is now known that these inhomogeneities in RGB stars are caused by two phenomena: multiple populations of stars in GCs and non-canonical, deep mixing in late-stage RGB stars \citep[][and references therein]{kraft94,gratton12}. Spectroscopic and photometric investigations in the last two decades have provided strong evidence of the presence of multiple stellar populations in all the GCs studied \citep[see e.g. the recent reviews by][and references therein]{bastian2018,gratton19}.

Notably, spectroscopic studies have demonstrated the existence on C-N, Na-O, and Mg-Al anti-correlations in all evolutionary stages in all Galactic GCs that have been studied, which indicates the existence of multiple populations of stars with different light element abundances \citep[see e.g.][and references therein]{gratton12}. More recent studies using Hubble Space Telescope (HST) UV photometry have also indicated the presence of multiple populations in every GC observed \citep{piotto,miloneatlas}. In the majority of GCs, these populations differ only in light element abundances which indicates that the progenitors that provided the intracluster material to form the additional populations did not enhance the heavy elements such as Fe.\footnote{About 15 percent of the GCs studied show even more complex chemical properties including variations in heavy elements; some of these GCs may be the cores of tidally disrupted dwarf galaxies accreted and, more in general, the formation history of these systems must include the additional contribution of progenitors enriching the cluster with heavy elements; see e.g. \citealt{johnson15b,marino15,yong15,dacosta16}.} This observation leaves only a few possible scenarios to create multiple populations, which are still heavily debated. Current candidates for the progenitor stars include massive asymptotic giant branch (AGB) stars \citep[see e.g.][]{ventura2001,dercole2008,dercole12}, fast rotating massive stars (FRMS) \citep{prantzos2006,decressin2007b}, massive binary stars \citep{demink2009}, or super-massive stars \citep{denissenkov2014,elmegreen17,gieles18}. All of these candidates would eject material slowly enough to be captured by the gravitational potential of the GC, and go through nucleosynthesis processes that could explain the anti-correlations observed such as the CN(O)-cycle and Ne-Na cycle.

In addition to initial differences in C and N caused by multiple populations, low-mass, low-metallicity RGB stars in GCs have been observed to go through an evolutionary process that causes additional C depletion and N enhancement to their surface abundances. This process, often referred to as ``extra" or ``secondary" mixing, takes place in RGB stars after they have passed a point on the color-magnitude diagram (CMD) called the luminosity function bump (LFB). The LFB is a result of canonical stellar evolution in RGB stars. When the expanding hydrogen (H) burning shell reaches a chemical weight difference (the $\mu$-barrier), the shell is given an influx of H causing the star to burn bluer for a brief moment before equilibrium is re-established and the star once again continues its ascent up the RGB. As multiple RGB stars ascend, then descend, then ascend once again, the result is a ``bump" in the luminosity function caused by multiple stars passing one another on the same point in the CMD. After RGB stars have passed this evolutionary stage, the $\mu$-barrier no longer prevents material from the H-burning envelope transferring into the outer convective envelope and being transported to the surface. Because this material has gone through the CN(O)-cycle, it is depleted in C, enhanced in N, and enhanced in C$^{13}$, which causes the observed changes in surface abundances of C, N, and $\Cratio$. However, the exact mechanism that causes this material to be transported between the H-burning envelope and the convective envelope is still not very well understood. Some of the main candidates include thermohaline mixing \citep{kippenhahn80,eggleton2006,eggleton2008,charbonnel07,henkel17}, rotational instability driven mixing \citep{sweigart79,chaname05,palacios06}, magnetic fields \citep{hubbard80,busso07,nordhaus08,palmerini09}, and internal gravity waves \citep{denissenkov00}, but no theory is currently able to match all observations without additional corrections.

For the following study, we focused on providing new observations to better understand these two phenomena through C and N abundances in the RGB stars of the low-metallicity GC, M53 (NGC 5024). Because both primordial chemical differences in the multiple populations present in GCs and secondary mixing will affect the surface abundances through CN(O)-cycle material, it can be difficult to separate the two effects when studying C and N in RGB stars. In order to disentangle this degeneracy in the possible origin of the variations in the surface abundances, we have observed a large number of stars ($\sim$100) covering a wide range of magnitudes from the tip of the RGB down to just below the LFB. We determine C and N abundances for the stars in our sample using low-resolution spectroscopy to measure CN and CH molecular features in the near-UV. We then use the strengths of these features and the C and N abundances to classify stars into different populations, and measure the change in surface abundances as a function of magnitude.

This work follows our similar studies of M10 \citep[][hereafter G18]{gerber18} and M71 \citep[][hereafter G20]{gerber20}. M53 is the low-metallicity cluster in this series with an [Fe/H] of -2.07 dex \citep{bobergm53}. Normally, CN bands are too weak to detect multiple populations in clusters of such a low metallicity, yet a previous study of a small sample of RGB stars in M53 found that the width of the distribution in CN band strengths exceeded the uncertainties, which indicated the presence of multiple populations \citep{martellmixing}. M53 also has Na and O abundance measurements from \citet{bobergm53}, which can be used to make comparisons between classifications based on N, Na, and O. We can also make comparisons to the metal poor clusters M92 and M15, which have relatively large sample sizes of C abundance measurements in the literature \citep[see][]{m3cfe,trefzger83}. 

Results from recent studies also highlight why M53 and its unusually nearby neighbor cluster, NGC 5053, are an important and unique pair of GCs to study. A potential tidal stream associated with NGC 5053 was observed by \citet{lauchner06}, which was later confirmed by \citet{jordi10}. \citet{chun2010} found evidence of a tidal bridge-like feature through wide field photometry between M53 and NGC 5053; however, this finding was not confirmed by \citet{jordi10}. \citet{forbes10} also suggested that M53 or NGC 5053 may be the remnant nucleus of a possible dwarf galaxy, although they mentioned more data such as proper motions were needed to confirm this suggestion. Since these results, proper motions from Gaia have revealed that the two clusters are likely on similar orbital trajectories, which provided more evidence that this pair may be an interacting pair of clusters \citep{vasiliev2019}. In addition, a recent study by \citet{yuan20} has found evidence of a stream associated with M53 and NGC 5053 (see also \citealt{naidu20}) and suggested that M53 may be the core of a disrupted dwarf galaxy, which would provide an explanation as to how a pair of clusters with a small distance from each other formed so far away from the Galactic center. Further evidence suggesting that M53 and NGC 5053 are a pair of clusters associated with the remnants of an accreted dwarf galaxy has also been presented by \citet{chun2020}. These recent findings make it even more important to study M53 and better understand its overall formation history.

The C and N abundances of M53 have not been well studied previously; the analysis of 85 RGB stars and 9 AGB stars presented in this paper provides a significant extension in the study of the chemical properties of this cluster. In Section \ref{observations}, we describe our observations and data reduction methods. We discuss the band measurements and calculations of C and N abundances for all stars in the sample in Section \ref{Analysis-m53}. Section \ref{Results-m53} discusses the relative proportions of first and second generation stars in our sample, the spatial distribution for the separate populations, a comparison with other identification methods, the evolution of C and N with magnitude, and a comparison to other clusters of similar metallicity (namely, M92 and M15). Our results and conclusions are summarized in Section \ref{Conclusions-m53}.

\section{Observations} \label{observations}

\subsection{Target Selection and Data Reduction} \label{targ-select}
We completed our observations over 3 observing runs on the WIYN\footnote{The WIYN Observatory is a joint facility of the University of Wisconsin-Madison, Indiana University, the National Optical-Infrared Astronomy Research Laboratory, the University of Missouri, Purdue University, Penn State University, and the University of California, Irvine.} 3.5m telescope during 9-12 Jun. 2016, 29 May - 1 Jun. 2017, and 22-24 Apr. 2018. We used the blue fibers of the Hydra multi-object spectrograph, which allows us to observe the near-UV CN band at 3883 \AA~and the CH band at 4300 \AA.

To acquire spectra of many stars with similar S/N using Hydra, it is best to observe stars in a configuration with a luminosity range of no more than 3 magnitudes. For this reason, our configurations for observations of M53 were divided into ``bright" and ``faint" set-ups. Our bright set-ups ranged in magnitude from 13.5 $<$ V $<$ 16, and our faint set-ups from 15.5 $<$ V $<$ 17.5. This allowed us to observe stars as faint as the LFB at V = 16.49 \citep{lfb}, which is the crucial point where extra mixing begins in RGB stars. The magnitude range of our sample can be seen in Figure \ref{fig::cmd-m53}, which is a  color-magnitude diagram (CMD) of M53 using photometry from \citet{rey98}.

While selecting stars for our sample, we focused on observing stars with Na and O abundances from 
\citet{bobergm53}. Observing stars with measured Na and O abundances provides us with the ability to compare our classification method using CN band measurements and the method using the Na-O anti-correlation \citep[see e.g.][]{gir,uves}. These comparisons are important as they can provide constraints on the progenitors for possible scenarios that lead to the formation of multiple populations. The O abundances also allow us to determine C and N abundances with higher accuracy as the balance between CN and CO molecular formation in a stellar atmosphere has a significant effect on the strength of the observed CN molecular feature.

Once observations were complete, our observed spectra were reduced using the IRAF\footnote{IRAF is distributed by the 
National Optical-Infrared Astronomy Research Laboratory, which is operated by the Association 
of Universities for Research in Astronomy (AURA) under cooperative agreement 
with the National Science Foundation.} package, \textit{dohydra}. With \textit{dohydra}, a one-dimensional spectrum is produced by tracing and extracting the light from each of the apertures from bias-subtracted images. Then the extracted spectra are flat-fielded, dispersion corrected, and sky-subtracted. Observations of a CuAr comparison lamp were used for dispersion correction. We then median-combined multiple observations of the same configuration using the IRAF package, \textit{scombine}, to increase the S/N and reduce the effects of cosmic rays.

We used the ``600@10.1" grating, which resulted in final spectra with a wavelength range of $\sim$3750-6300 \AA~and a FWHM of $\sim$4.5 \AA. This telescope set up and reduction procedure is similar to what was used in G18 and G20 to collect the datasets in M10 and M71. The minimum S/N at the position of the CN band at 3883 \AA~is 20 and can be as high as 50 in some spectra with a typical S/N being $\sim$30. The S/N at the position of the CH band (4300 \AA) is much higher with the lowest value observed being 100 and some spectra having a S/N as high as 200. A typical S/N value is $\sim$150.

\subsection{Radial Velocities and Membership Selection}

We initially selected likely members to observe based on their location on the cluster CMD. Once observations were complete, we then used radial velocity (RV) measurements to further ensure that the stars were cluster members. RV measurements were made by cross-correlating absorption lines in the observed stellar spectra with a spectrum of a RV standard using the IRAF package, \textit{fxcor}. We chose to use HD107328, HD109358, and HD132737 as RV standards as all three stars have a similar temperature and luminosity as the program stars. Observations of the three RV standards were made during each observing run. Uncertainties for individual measurements using this method at this resolution are around 15 km s$^{-1}$.

\begin{figure}[htbp]
\centering
\includegraphics[trim = 0.2cm 0.4cm 0.4cm 0.4cm, scale=0.4, clip=True]{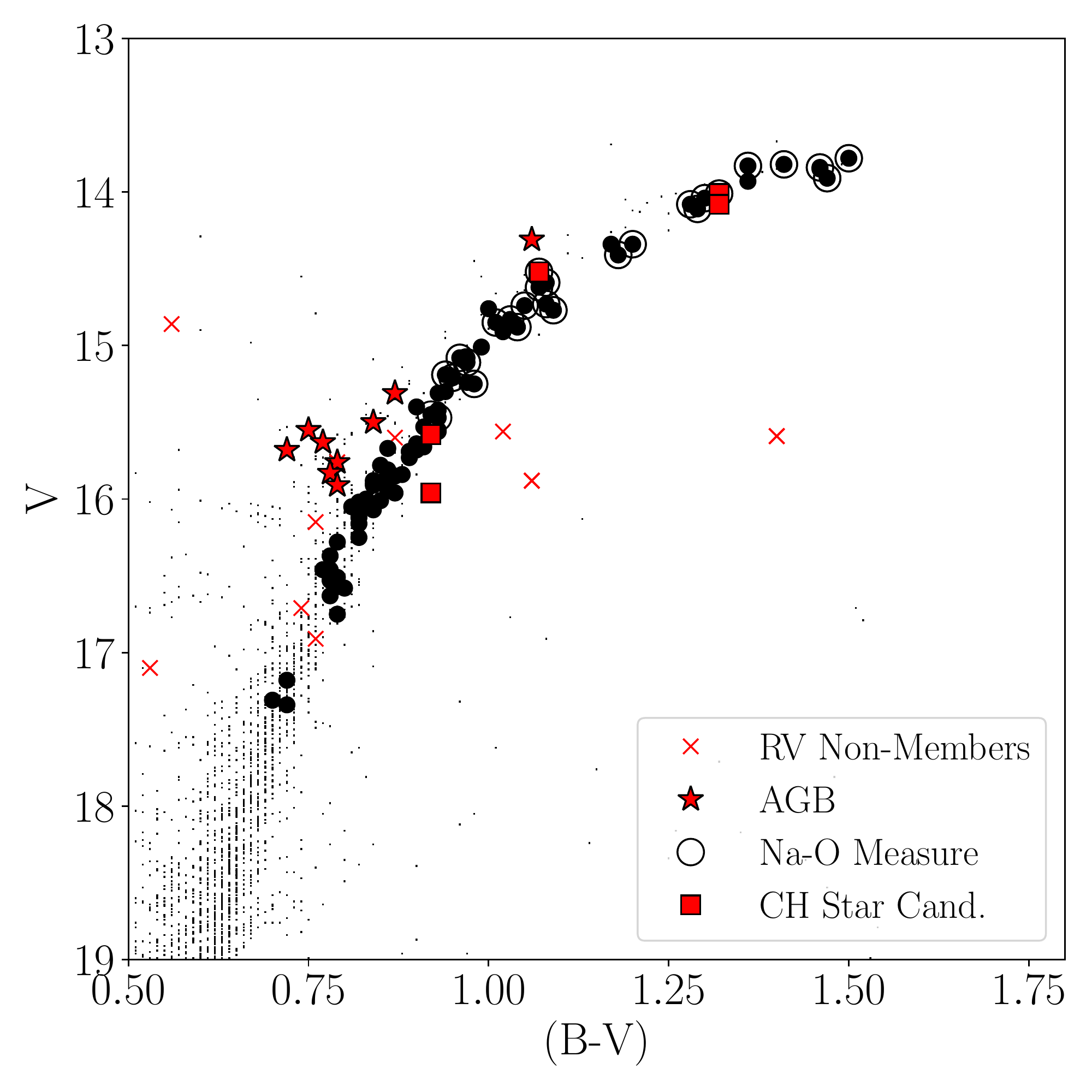}
\caption{CMD of M53 using B and V photometry from \citet{rey98}. Black circles and red stars indicate observed RGB and AGB member stars, respectively. Red x's indicate stars that were determined to be non-members based on their RVs. Red squares are CH star candidates as discussed in Section \ref{CH-cands}. Stars with Na or O abundance measurements from \citet{bobergm53} are indicated.}
\label{fig::cmd-m53}
\end{figure}

Our initial sample included 148 individual spectra. Forty-one measurements of multiple stars were included in the M53 sample, and these measurements were used to check for any offsets between nights or observing runs; none were found. To constrain membership, any stars with RVs above or below three standard deviations of the median were determined to be non-members and removed from the sample. This limit resulted in 9 stars being classified as non-members based on RV.
The final sample for M53 includes 85 RGB and nine AGB stars; these are shown in Figure \ref{fig::cmd-m53} along with non-members and stars with Na or O abundances. We find a median RV of -60 $\pm$ 9 (std) km s$^{-1}$, which agrees with the value found by \citet{bobergm53} (-63.2 $\pm$ 0.5 km s$^{-1}$) and \citet{kimmig15} (-62.8 $\pm$ 0.3 km s$^{-1}$) from higher resolution data. A full list of objects observed, including RV non-members, is presented in Table \ref{tab:alldata}.

\section{Analysis} \label{Analysis-m53}

\subsection{CN and CH Bands}

\subsubsection{Index Definitions}

To measure the CN and CH molecular features, we made use of indices that create a ``magnitude" of band strength by comparing the integrated flux of a feature bandpass to one or more continuum bandpasses. CN and CH indices have been used in many studies to identify multiple populations and measure C and N abundances in stars in GCs \citep[e.g.][]{norris79,bands,briley2004b,smolinski11}.

Our index definitions were chosen such that they would be accurate for the large range of magnitudes (meaning large ranges in effective temperatures and surface gravities) that our sample covers. As in G18 and G20, we use the S(3839) and CH(4300) (referred to as CH throughout the rest of this paper) bands defined by \citet{bands}. This S(3839) index is based on the S(3839) index defined by \citet{norris81} with the difference being slightly tighter feature and continuum bandpass windows to avoid contamination from stronger Balmer lines in the fainter/warmer stars. The CH index is based on a similar index used by \citet{cohen99a,cohen99b} with continuum bandpasses chosen to avoid strong absorption by H$\gamma$ in warmer stars. Our final band definitions are:

\begin{equation}
S(3839) = - 2.5log\frac{F_{3861-3884}}{F_{3894-3910}}
\end{equation}
\begin{equation}
CH = -2.5log\frac{F_{4285-4315}/30}{0.5(F_{4240-4280}/40) + 0.5(F_{4390-4460}/70)}
\end{equation}

\noindent where $F_{x-y}$ is the integrated flux between the wavelengths x and y in \AA.

\subsubsection{Flux Calibration} \label{fluxcal}
We used the same method described in G18 and G20 to flux calibrate the spectra using model spectra generated by the Synthetic Spectrum Generator (SSG) \citep[][and references therein]{ssg} using MARCS model atmospheres \citep{gustafsson08}. We used the average cluster metallicity of [Fe/H] = -2.07 from \citet{bobergm53}, and calculated effective temperatures from the V-K colors of the stars based on the method by \citet{alonso99, alonso01} with V magnitudes from \citet{rey98} and K magnitudes from 2MASS \citep{2mass}. K magnitudes were corrected to the Carlos S{\'a}nchez Telescope (TCS) system \citep{tcs} following the method by \citet{johnson05}. If a star did not have a K magnitude, the relation for effective temperature based on B-V color was used \citep{alonso99}. To correct for extinction, a reddening correction of E(B-V) = 0.02 \citep{rey98} was used for all stars in the cluster. 

Surface gravities were calculated using the bolometric corrections given by \citet{alonso99} and are listed with the effective temperature for each star in Table 1. Absolute magnitudes were calculated for each star using an apparent distance modulus of (m-M)$_\mathrm{V}$ = 16.32 (\citealt{harris}: 2010 edition). These are also given in Table 1.

Using these stellar parameters, we created a scaled-solar spectrum for each star in the sample using the SSG. Each observed spectrum was divided by a corresponding scaled-solar spectrum matching its stellar parameters. The resulting ratio was then fit with a spline to create a function fitting the approximate instrumental response for a given spectrum. Each observed spectrum was then divided by the instrumental response function of that star to generate our sample of spectra with continua that had been calibrated to the models. For more details on this method see G18.

\subsubsection{Band Measurements} \label{bandmeasurements-m53}
 
\begin{figure}
\centering
\includegraphics[trim = 0.0cm 0.0cm 0.0cm 0.0cm, scale=0.4, clip=True]{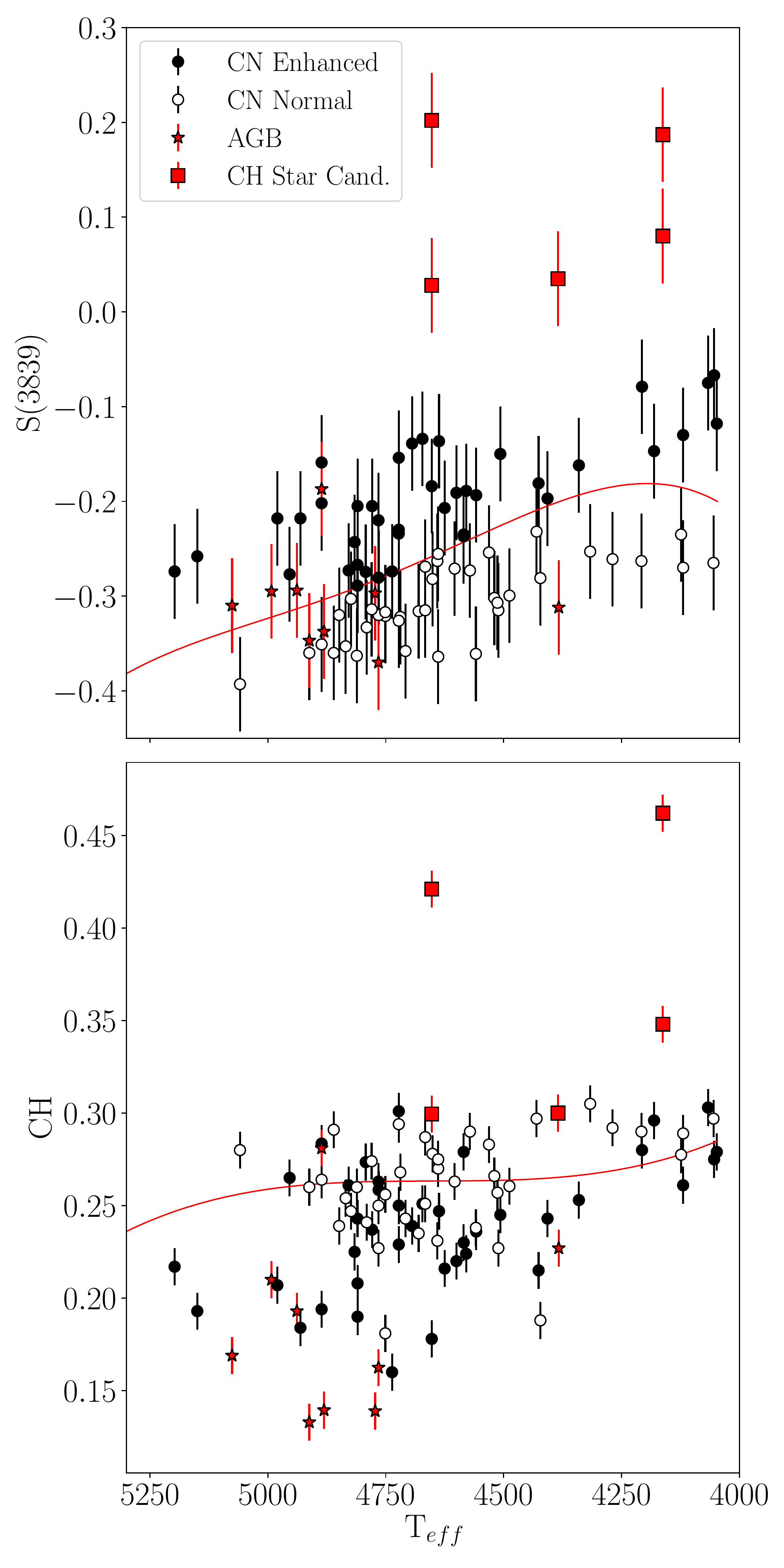}
\caption{\textbf{Top Panel:} CN band strengths as a function of effective temperature for stars in our sample for M53. The red line shows the fiducial used to determine the $\delta$CN index strength, and identify a star as CN-enhanced (filled) or CN-normal (open). AGB stars are indicated as red stars, RGB stars are shown as circles (filled or open), and CH star candidates are shown as red squares. \textbf{Bottom Panel:} CH band strengths as a function of effective temperature using the same symbols as the upper panel. The same fiducial calculated for CH band strength is also shown. 
Both bands show a clear trend in temperature.}
\label{bands-m53}
\end{figure}

We measured S(3839) and CH index values using the flux-calibrated spectra for all of the stars in our sample for M53. These measurements are plotted versus effective temperature in Figure \ref{bands-m53} and listed in Table \ref{tab:alldata}. Error bars in Figure \ref{bands-m53} represent the calculated uncertainties for each band. Using multiple measurements of the same stars, we found that the S(3839) and CH band indices for stars in our sample have uncertainties of 0.05 and 0.01, respectively.

Figure \ref{bands-m53} shows that both bands increase in strength as a function of decreasing effective temperature. While there is a spread in the S(3839) index measurements indicative of more than one population, the S(3839) index does not clearly separate into two populations except at the coolest temperatures, which is likely due to the extremely low metallicity of M53. Since the stars in this cluster are so metal poor, the CN bands are weaker in their spectra, which reduces any visible differences between populations that may exist. Therefore, to better quantify whether a star is CN-enhanced or CN-normal, we used an objective classification method.

As in G18 and G20, our method involves defining a $\delta$CN index by modeling and removing the changes in CN band strength as a function of effective temperature from our S(3839) index measurements. We model these changes using C and N abundance measurements made using our S(3839) and CH index measurements. To measure C and N in our stars, we use the SSG following the methods of \citet{briley2004a,briley2004b}. Synthetic spectra are generated for each star with the effective temperature and surface gravity that was determined in Section \ref{fluxcal} and given in Table \ref{tab:alldata}, and an assumed initial C and N abundance. Then, the S(3839) and CH indices are measured in the synthetic spectrum, and the C and N abundances are simultaneously varied until the synthetic S(3839) and CH indices match what is observed.

\begin{deluxetable*}{cccccccccccccccccc}
\tabletypesize{\scriptsize}
\tablecolumns{18}
\tablewidth{0pt}
\tablecaption{Stars Measured in M53
\label{tab:alldata}}
\tablehead{\\
\colhead{ID$^{1}$} & \colhead{RA$^{1}$} & \colhead{Dec$^{1}$} & \colhead{RV$_{hel}$} & \colhead{$V^{1}$} & \colhead{M$_\mathrm{V}^{2}$} & \colhead{$K_{2MASS}$} & \colhead{$T_{eff}$} & \colhead{logg} & \colhead{S(3839)} & \colhead{$\delta$CN} & \colhead{CH} & \colhead{[C/Fe]} & \colhead{[N/Fe]} & \colhead{[O/Fe]$^{3}$} & \colhead{[Na/Fe]$^{3}$} & \colhead{Branch} & \colhead{Memb} \\ 
& \colhead{(J2000)} & \colhead{(J2000)} & \colhead{(km s$^{-1}$)} & (mag) & (mag) & (mag) & (K) & (cm s$^{-1}$) & & & & & & & &}
\rotate
\startdata
    4 &  198.191361 &  18.191845 &   -54 &  13.78 & -2.54 &  10.441 &  4048 &  0.68 &   -0.118 &       0.082 &     0.279 &   -0.68 &    1.71 &      0.4 &     0.5 &       RGB &    y \\
    7 &  198.238318 &  18.149578 &   -60 &  13.82 & -2.50 &  13.640 &  4067 &  0.71 &   -0.075 &       0.120 &     0.303 &   -0.57 &    1.58 &      0.1 &     0.4 &       RGB &    y \\
    9 &  198.253564 &  18.167292 &   -60 &  13.83 & -2.49 &  13.445 &  4119 &  0.76 &   -0.270 &      -0.084 &     0.289 &   -0.61 &    1.13 &     -0.1 &     0.4 &       RGB &    y \\
   10 &  198.232289 &  18.189502 &   -54 &  13.84 & -2.48 &  10.514 &  4055 &  0.71 &   -0.265 &      -0.067 &     0.297 &   -0.57 &    1.20 &     -0.2 &     0.5 &       RGB &    y \\
   12 &  198.233857 &  18.139113 &   -56 &  13.85 & -2.47 &  10.522 &  4054 &  0.71 &   -0.067 &       0.131 &     0.275 &   -0.75 &    1.80 & \nodata & \nodata &        RGB &    y \\
   14 &  198.268107 &  18.182968 &   -53 &  13.91 & -2.41 &  10.707 &  4124 &  0.79 &   -0.235 &      -0.050 &     0.278 &   -0.51 &    1.35 &      0.0 &     0.7 &       RGB &    y \\
   15 &  198.211288 &  18.161073 &   -53 &  13.93 & -2.39 &  10.719 &  4119 &  0.80 &   -0.130 &       0.056 &     0.261 &   -0.77 &    1.64 & \nodata & \nodata &        RGB &    y \\
   18 &  198.210175 &  18.178769 &   -53 &  14.01 & -2.31 &  13.039 &  4162 &  0.86 &    0.080 &       0.262 &     0.348 &   -0.31 &    1.55 &      0.2 &     0.4 &  CH Star Cand. &    y \\
   20 &  198.235309 &  18.144363 &   -56 &  14.04 & -2.28 &  10.974 &  4208 &  0.91 &   -0.263 &      -0.082 &     0.290 &   -0.46 &    1.07 &     -0.2 &     0.5 &       RGB &    y \\
   26 &  198.251358 &  18.168852 &   -72 &  14.08 & -2.24 &  14.115 &  4162 &  0.89 &    0.187 &       0.369 &     0.462 &    0.13 &    1.18 & \nodata & \nodata &   CH Star Cand. &    y \\
   28 &  198.218386 &  18.173816 &   -57 &  14.08 & -2.24 &  12.446 &  4207 &  0.92 &   -0.079 &       0.102 &     0.280 &   -0.61 &    1.52 &      0.3 &     0.3 &       RGB &    y \\
   29 &  198.209281 &  18.149267 &   -59 &  14.11 & -2.21 &  11.001 &  4181 &  0.92 &   -0.147 &       0.034 &     0.296 &   -0.47 &    1.36 &     -0.2 &     0.5 &       RGB &    y \\
   38 &  198.261212 &  18.184932 &   -61 &  14.31 & -2.01 &  11.497 &  4383 &  1.13 &   -0.312 &      -0.112 &     0.227 &   -0.63 &    1.16 & \nodata & \nodata &        AGB &    y \\
   39 &  198.198274 &  18.169346 &   -77 &  14.34 & -1.98 &  11.469 &  4340 &  1.12 &   -0.162 &       0.031 &     0.253 &   -0.59 &    1.41 & \nodata & \nodata &        RGB &    y \\
   40 &  198.203296 &  18.238325 &   -64 &  14.34 & -1.98 &  11.367 &  4269 &  1.07 &   -0.261 &      -0.076 &     0.292 &   -0.41 &    1.05 &      0.0 &     0.5 &       RGB &    y \\
   44 &  198.183393 &  18.170152 &   -66 &  14.41 & -1.91 &  11.505 &  4317 &  1.13 &   -0.253 &      -0.063 &     0.305 &   -0.32 &    0.99 &     -0.2 &     0.5 &       RGB &    y \\
   47 &  198.233739 &  18.185220 &   -64 &  14.52 & -1.80 &  11.709 &  4385 &  1.22 &    0.035 &       0.236 &     0.300 &   -0.43 &    1.51 &      0.3 &     0.1 &  CH Star Cand. &    y \\
   53 &  198.201814 &  18.213504 &   -64 &  14.59 & -1.73 &  11.808 &  4407 &  1.26 &   -0.197 &       0.008 &     0.243 &   -0.58 &    1.35 &      0.0 &     0.3 &       RGB &    y \\
   54 &  198.270264 &  18.247453 &   -62 &  14.62 & -1.70 &  11.940 &  4488 &  1.32 &   -0.300 &      -0.078 &     0.260 &   -0.35 &    1.02 &      0.0 &     0.5 &       RGB &    y \\
   60 &  198.236786 &  18.100674 &   -56 &  14.73 & -1.59 &  12.073 &  4507 &  1.38 &   -0.150 &       0.076 &     0.245 &   -0.47 &    1.42 &      0.3 &     0.2 &       RGB &    y \\
   61 &  198.231885 &  18.220191 &   -67 &  14.74 & -1.58 &  12.111 &  4531 &  1.40 &   -0.254 &      -0.023 &     0.283 &   -0.24 &    1.06 &      0.0 &     0.4 &       RGB &    y \\
   64 &  198.253228 &  18.200100 &   -64 &  14.76 & -1.56 &  11.997 &  4422 &  1.34 &   -0.281 &      -0.073 &     0.188 &   -0.80 &    1.48 & \nodata & \nodata &        RGB &    y \\
   65 &  198.152838 &  18.126203 &   -60 &  14.77 & -1.55 &  13.076 &  4430 &  1.35 &   -0.232 &      -0.023 &     0.297 &   -0.30 &    1.05 &     -0.1 &     0.4 &       RGB &    y \\
   68 &  198.259119 &  18.222985 &   -58 &  14.83 & -1.49 &  12.188 &  4520 &  1.43 &   -0.302 &      -0.074 &     0.266 &   -0.29 &    1.02 &      0.0 &     0.5 &       RGB &    y \\
   70 &  198.201253 &  18.238394 &   -65 &  14.85 & -1.47 &  12.198 &  4511 &  1.43 &   -0.315 &      -0.088 &     0.227 &   -0.53 &    1.12 &      0.0 &     0.3 &       RGB &    y \\
   72 &  198.199614 &  18.109783 &   -59 &  14.88 & -1.44 &  12.230 &  4513 &  1.44 &   -0.307 &      -0.080 &     0.257 &   -0.40 &    1.01 &      0.0 &     0.3 &       RGB &    y \\
   77 &  198.262790 &  18.195211 &   -62 &  14.91 & -1.41 &  12.153 &  4426 &  1.40 &   -0.181 &       0.028 &     0.215 &   -0.68 &    1.55 & \nodata & \nodata &        RGB &    y \\
   84 &  198.287056 &  18.188487 &   -69 &  15.01 & -1.31 & \nodata &  4558 &  1.52 &   -0.194 &       0.044 &     0.236 &   -0.40 &    1.39 & \nodata & \nodata &        RGB &    y \\
   86 &  198.227725 &  18.092297 &   -58 &  15.08 & -1.24 &  12.615 &  4672 &  1.61 &   -0.134 &       0.129 &     0.251 &   -0.16 &    1.50 &      0.0 &     0.5 &       RGB &    y \\
   87 &  198.208583 &  18.197453 &   -60 &  15.08 & -1.24 &  12.484 &  4559 &  1.55 &   -0.361 &      -0.124 &     0.238 &   -0.37 &    0.92 & \nodata & \nodata &        RGB &    y \\
   91 &  198.207839 &  18.187731 &   -74 &  15.11 & -1.21 &  15.192 &  4585 &  1.57 &   -0.235 &       0.008 &     0.230 &   -0.39 &    1.34 &      0.1 &     0.4 &       RGB &    y \\
   95 &  198.241690 &  18.153194 &   -59 &  15.19 & -1.13 &  14.510 &  4625 &  1.63 &   -0.207 &       0.045 &     0.216 &   -0.42 &    1.45 &      0.2 &     0.2 &       RGB &    y \\
   97 &  198.235564 &  18.210077 &   -62 &  15.21 & -1.11 &  12.653 &  4579 &  1.61 &   -0.189 &       0.053 &     0.224 &   -0.48 &    1.44 &      0.0 &     0.1 &       RGB &    y \\
   99 &  198.264958 &  18.148353 &   -58 &  15.24 & -1.08 &  15.069 &  4585 &  1.63 &   -0.237 &       0.006 &     0.279 &   -0.21 &    1.14 & \nodata & \nodata &        RGB &    y \\
  101 &  198.149573 &  18.241642 &   -60 &  15.25 & -1.07 & \nodata &  4571 &  1.62 &   -0.273 &      -0.033 &     0.290 &   -0.17 &    1.00 &      0.1 &     0.4 &       RGB &    y \\
\enddata
\tablecomments{1. \citet{rey98}, 2. Assumes (m -- M)$_{v}$ = 16.32 \citealt{harris} (2010 edition), 3. \citet{bobergm53} \\}
\end{deluxetable*}

Other than initial C and N abundances, we also assumed a microturbulence of 2.0 km s$^{-1}$ and a $C^{12}/C^{13}$ ratio of 4.0, which are representative of RGB stars in GCs \citep[e.g.,][]{suntzeff91,pavlenko}. For accurate models, we also have to assign an oxygen abundance for each star due to the dependence of CN and CH band strength on CO molecular formation in a star's atmosphere. 
We use an average [O/Fe] value for the cluster of 0.4 dex based on the measurements from \citet{bobergm53}. This assumption works for the purposes of these preliminary measurements as the main goal is to determine the average C and N abundances as a function of luminosity.

The stellar C and N abundances were used to determine the average C and N abundance at 100 K intervals. Using these average abundances, we created average S(3839) and CH index measurements over the same intervals by measuring synthetic spectra created with the SSG. A spline was then fitted to these points, and used as a dividing line between the two populations; see the red lines in Figure \ref{bands-m53}. These lines represent the average S(3839) or CH index measurement for a given effective temperature. The $\delta$CN index was calculated by subtracting the fit to the average at that effective temperature from the measured S(3839) value. Our definitions for each population match those from G18 and G20, with CN-enhanced stars having $\delta$CN $>$ 0.0 and CN-normal stars having $\delta$CN $\leq$ 0.0. In Figure \ref{bands-m53}, CN-enhanced stars are indicated as filled points and CN-normal stars as open points.

We also identify in Figure \ref{bands-m53} five stars with abnormally high S(3839) measurements; all also have enhanced CH strengths. The $\delta$CN index measured for these stars is almost twice as high as the next highest value of a CN-enhanced star, suggesting they have special properties. We defer a discussion of these stars, which we identify as potential CH stars, until Section \ref{CH-cands}, but label them as red squares in all figures so their behavior within the sample can be distinguished. We also present, for completeness, our calculated abundance values for each of these CH star candidates, and include them with the caveat that our analysis technique may not be capable of producing truly accurate values for stars with such extreme CN and CH molecular features.

\subsection{Determining C and N Abundances} \label{candn-m53}
To determine our final C and N abundances for each star, we follow the same method described above to find initial abundances. The only difference in our final abundances is that once the stars are classified into populations, more specific [O/Fe] values for each star are used. We adopted the [O/Fe] value from \citet{bobergm53} for the 23 stars with measurements. If [O/Fe] was not measured for the star, we assumed an [O/Fe] of 0.3 dex for CN-enhanced stars and 0.47 dex for CN-normal stars, which are the average values for each population based on the O abundances from \citet{bobergm53}. This method provides more accurate results than assuming a single average [O/Fe] value for the entire sample because \citet{bobergm53} shows a 0.6 dex range in O abundances in the cluster. Making an accurate assumption for the [O/Fe] value is important because differences in O abundances of 0.3 dex can cause changes in the final C and N abundances of up to 0.2 dex in fainter, warmer stars.

The uncertainties in these abundance determinations were calculated using the same method outlined in G18. We adjust the temperature by 150 K, change the [O/Fe] of the CN-enhanced and CN-normal stars, and increase and decrease the S(3839) and CH bands, respectively, by their uncertainties. We add the resulting changes to [C/Fe] and [N/Fe] in quadrature and find uncertainties of 0.2-0.25 dex for each abundance for stars with M$_\mathrm{V} \leq 0$. Fainter stars are found to have slightly higher uncertainties of 0.35 dex, which is mostly due to the CN bands' loss of sensitivity to changes in abundance. A more detailed discussion on our uncertainties using this method can be found in G18. 

\begin{figure}
\centering
\includegraphics[trim = 0.4cm 0.4cm 0.4cm 0.4cm, scale=0.34, clip=True]{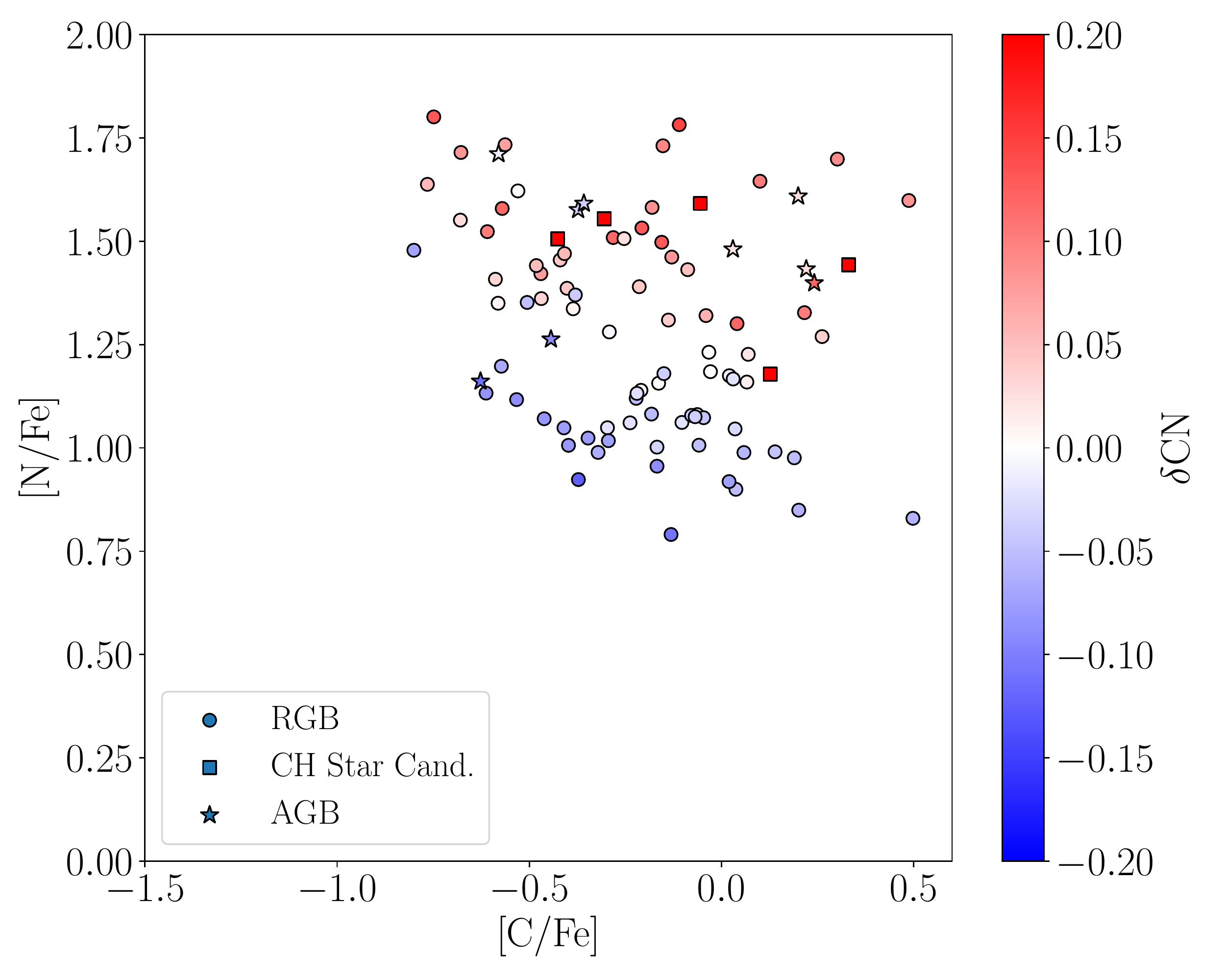}
\caption{[N/Fe] vs. [C/Fe] for all of the members in our sample in M53. Following the same convention as previous figures, AGB stars are indicated as stars, and the candidate CH stars from Figure \ref{bands-m53} are indicated as squares. Points are color-coded based on their $\delta$CN index as indicated by the color bar on the right side of the plot.}
\label{nfevscfe-m53}
\end{figure}

Figure \ref{nfevscfe-m53} shows our final derived [C/Fe] and [N/Fe] measurements for each star color-coded by their $\delta$CN values. As expected, we find a spread in C and N with values in the abundance of each element having a range of 1.0 dex or more. This result is similar to what G18 observed in M10, although for M53, the [N/Fe] values are slightly higher than those measured for M10. CN-enhanced and CN-normal RGB stars also map directly onto enhanced and normal N abundances, respectively. Unlike what was found in M10 by G18, Figure \ref{nfevscfe-m53} shows that the RGB stars in M53 do not separate cleanly into two populations in the C-N plane, but rather show a large range of C and N abundances in each population. Further comparison with M10 shows that while the range in [C/Fe] in the two clusters is roughly the same, the range in [N/Fe] in M53 (1 dex) is smaller than that observed in M10 ($\sim$2 dex).

The distribution in Figure \ref{nfevscfe-m53} and large range in abundances is influenced strongly by evolutionary processes. Because of M53's distance and therefore, relatively fainter apparent magnitude compared to other clusters, our sample consists of stars that are mostly above the LFB (M$_\mathrm{V}$ = 0.168; \citealt{lfb}). Since almost all stars in our sample are brighter than the LFB, they have an underlying dependence of abundance on stellar magnitude, with brighter stars having lower C and enhanced N. This effect works to smooth any initial differences in abundance between the CN-enhanced and CN-normal populations that would initially show two distinct distributions in the C-N plane. We discuss these evolutionary effects further in Section \ref{evolution-m53}.

\section{Results and Discussion} \label{Results-m53}

\subsection{Multiple CN Populations in M53}

\begin{figure}
\centering
\includegraphics[trim = 0.4cm 0.4cm 0.4cm 0.4cm, scale=0.38, clip=True]{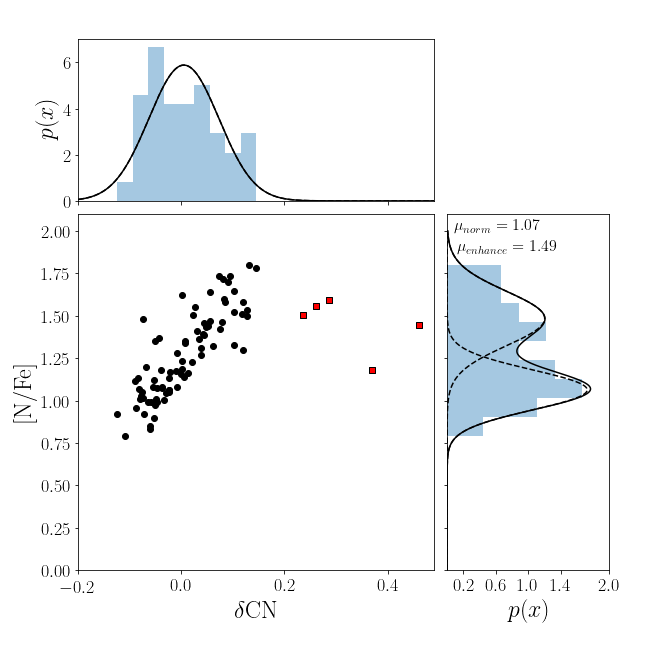}
\caption{[N/Fe] vs. $\delta$CN for the RGB stars in the sample from M53. CH star candidates are indicated as red squares. Probability distributions for the $\delta$CN and [N/Fe] are shown on the x and y axis, respectively. Data are shown in blue with the two Gaussian distributions yielded by GMM overplotted. The mean values of the fit to the [N/Fe] distribution are given at the top of the histogram. $\delta$CN correlates, as expected, with [N/Fe], and the [N/Fe] distribution shows the clear presence of two distinct populations.}
\label{gmm-m53}
\end{figure}

In Figure \ref{gmm-m53}, we plot our $\delta$CN measurements vs. [N/Fe] for all of the RGB stars in our sample. As expected, $\delta$CN correlates with [N/Fe] for the stars in M53. The only exceptions to this correlation are five stars with anomolously high $\delta$CN measurements, which we discuss more fully in Section \ref{CH-cands}. These stars have relatively large $\delta$CN values compared to the rest of the stars in the cluster with all five having $\delta$CN $>$ 0.2. However, we do not find these stars to have significantly larger [N/Fe] measurements than what is seen in the [N/Fe] distribution for the rest of the sample. Because the CN molecular band is dependent on both N and C, the enhanced CN can be equally explained by a high C abundance, and since all of these stars have strong CH bands (as shown in Figure \ref{bands-m53}), they have relatively enhanced C abundances compared to other stars of similar magnitude (see G18 for a more detailed discussion of how an enhanced CN band does not necessarily imply an enhancement in the underlying N abundance).

The marginal distributions in $\delta$CN and [N/Fe] are also shown in Figure \ref{gmm-m53} along the x and y axes, respectively. While the $\delta$CN index distribution does not allow a clear separation of multiple populations, the [N/Fe] distribution shows two distinct peaks and confirms the presence of two stellar populations. This figure clearly illustrates the importance of the information from N abundance measurements for the identification of multiple populations in comparison with separation of populations based solely on a $\delta$CN index. 
Because M53 has such a low metallicity, the difference in $\delta$CN strength between the two populations is too small to easily distinguish two significant tracks in the S(3839) vs. T$_{eff}$ plot. However, the difference in [N/Fe] between each population becomes clearly distinguishable as two peaks in the distribution. These populations are producing the broad range in CN strength, which is larger than our observational error.

We fit the distribution of [N/Fe], including the stars with high $\delta$CN, with a Gaussian Mixture Model (GMM). To determine the number of populations to fit to the data, we fit one to ten Gaussians to the distribution and measure the Akaike Information Criterion (AIC) \citep{akaike74} and the Bayesian Information Criterion (BIC) \citep{schwarz78}. Using these criteria, we then apply a GMM with two populations as this number minimized both the AIC and BIC, indicating it was the best fit. The two populations in [N/Fe] are found to have a separation of 0.4 dex and similar standard deviations of 0.13 and 0.16 for the first and second generation, respectively. The separation between populations and the standard deviations do not change significantly when the stars shown with $\delta$CN values greater than 0.2 are excluded. Figure \ref{gmm-m53} shows the GMM fit to the [N/Fe] distribution. We also show the GMM fit to the $\delta$CN distribution, which does not include the stars with high $\delta$CN. The AIC and BIC indicate that the $\delta$CN distribution is best fit by a single population, which could indicate issues with using $\delta$CN to classify stars into two populations. However, because of the strong correlation between [N/Fe] and our $\delta$CN index, and since a $\delta$CN of 0.0 matches the separating value of 1.28 dex between the two Gaussian distributions for the [N/Fe] distribution, we are confident in our classification by $\delta$CN index.

We find that 40 out of 80 of RGB stars in our sample are CN-enhanced based on having a positive $\delta$CN measurement. We do not include in this group the 5 stars with $\delta$CN $>$ 0.2; these 5 stars are all classified as CN-enhanced based on their $\delta$CN index, but as seen in Figure \ref{gmm-m53}, these stars are not all N-enhanced. If we calculate the ratio of second generation stars based on [N/Fe], including all stars, we find that 42 out of 85 of stars are N-enhanced. A value of [N/Fe] equal to 1.28 dex, the midpoint between the two Gaussian curves fit to the [N/Fe] distribution, is adopted to separate the two populations.

\subsubsection{AGB Stars} \label{AGB-m53}
There has been some recent debate over whether or not both populations of stars in GCs evolve onto the AGB. For example, \citet{campbell12,campbell13} and \citet{maclean16} found an absence of the CN-enhanced population on the AGBs in NGC 6257 and M4. These studies suggested that the lack of these stars could be an indication that the He abundance in the second generation of stars is high enough to cause them to skip that phase of stellar evolution. Later studies, however, were able to observe both populations of stars in 47 Tuc, M13, M5, M3, M2, and NGC 6397 \citep{johnson15,garcia15,maclean18}, and a recent study by \citet{lagioia2021} observed both populations of stars in all of the clusters studied. A more detailed review of this debate can be found in G18.

In order to further explore this issue, we have observed AGB stars in M53. However, our final sample of AGB stars in M53 includes only 9 stars. Consequently, the conclusions concerning the presence of multiple populations in the AGB phase is not as strong as in our previous studies of M10 and M71 (G18; G20). We also note that our classification of these stars is not as secure as it was for M10 and M71 since the AGB of M53 is not as clearly defined in the CMD. We present our findings on AGB stars below with this disclaimer in mind.

Our band measurements for the AGB stars in M53 are shown in Figure \ref{bands-m53}. Based on the S(3839) measurements, the AGB stars in our sample seem to be evenly divided between the CN-enhanced and CN-normal populations with 4 out of 9 AGB stars being CN-enhanced. However, as we discussed in G18, the CN band is dependent on both the C and N abundance making it important to consider the CH band strength before concluding that a strong or weak CN band is caused solely by N. The lower panel of Figure \ref{bands-m53} shows that the AGB stars have relatively weak CH bands (or low C abundances), which will require relatively higher N abundances to explain the CN band strength. This effect explains how AGB stars in Figure \ref{nfevscfe-m53} with low $\delta$CN values can be found having high [N/Fe] values.

When separating populations by N abundance, we find that 7 out of 9 AGB stars are N-enhanced. While this ratio does not match that of the RGB stars, our uncertainty in the classification of AGB stars, and our relatively small sample of AGB stars, makes it difficult to make any firm conclusions about whether this difference is a physical result or simply a case of large uncertainties for a small sample. Using a larger sample of stars, \citet{lagioia2021} found that AGB stars in M53 had a ratio of first generation to second generation stars similar to the RGB. Although our study does not have a large enough sample to confirm these results, our analysis shows clear evidence that second generation (or CN-enhanced) stars are present on the AGB in M53.

\subsection{Candidate CH Stars} \label{CH-cands}

CH stars are a type of C-enhanced star found along the RGB, and are identified by an extreme level of C enhancement leading to the presence of unusually strong CH and C$_2$ molecular spectral features. Since these low-mass RGB stars are not expected to have gone through a third-dredge up, the cause of the enhanced C and s-process surface abundances also often seen was a mystery until it was discovered that field CH stars have a binary white dwarf companion \citep{mcclure90}. Mass transfer due to Roche lobe overflow or stellar winds between the low-mass RGB progenitor and the more massive white dwarf progenitor as it ascended the AGB was then thought to be the cause of these observed abundance peculiarities. 
This mechanism requires the survival of binaries with semi-major axes large enough to allow the mass transfer to occur during the AGB stage \citep[see e.g.][]{cote97}. However, recent studies have suggested that some species of CH stars with more moderate enhancement may be a result of incomplete CN processing rather than mass exchange \citep{sharina12}. It is clear that these objects require further study to understand how they are formed.

CH stars are rarely found in GCs; searches have identified possible CH stars in only $\omega$ Cen, M22, M55, M2, M14, M15, and NGC 6426 and these show a range of CH star features \citep{harding62,dickens72,cowley85,mcclure77,smith82,smith90,lardo12,cote97,kirby15,sharina12}. Classic CH stars with strong C$_2$ bands have only been found in $\omega$ Cen and M22 \citep{harding62,dickens72,cowley85,mcclure77}, but in clusters such as M14, M15, and NGC 6426, CH stars with more mild indicators such as strong CH but weak C$_2$ are observed \citep{cote97,kirby15,sharina12}. These differences could be explained by a different formation mechanism such as the incomplete CN processing mentioned above \citep{sharina12}, but might also be caused by warmer temperatures that would prevent the formation of strong C$_2$ bands in their spectra.

In addition to solving issues of formation mechanisms, observing such stars in GCs is important to shed light on the possible connection between the origin of CH stars and the survival and evolution of binary stars in the clusters' dense environments. In order to explore the link between the presence of CH stars and the dynamics of binary stars, \citet{cote97} calculated the characteristic value of the semi-major axis, $a_h$, of surviving binaries in the central regions of GCs; although, as they pointed out, their estimate can not capture the much more complex evolution of binaries in clusters, it is nevertheless interesting that for several of the clusters in which CH stars have been found, $a_h$ is larger than (or smaller but close to) the value needed for a binary to produce a CH star. Using the values of the central density and velocity dispersion estimated by \citet{mclaughlin05} for M53 we find that also for this cluster $a_h$ is larger ($a_h \sim 1.86$ AU and $a_h \sim 2.56$ AU using the structural parameters obtained by \citet{mclaughlin05}, respectively, with a King model and a Wilson model fit) than the critical value ($a_h \sim 1$ AU) adopted by \citet{cote97} to determine whether a cluster is likely to host CH stars produced from the evolution of binary stars. In this context, we discuss the identification of potential CH stars in M53.

As noted above, in addition to the two populations of stars separated by S(3839) values in Figure \ref{bands-m53}, we observed five stars with anomalously high S(3839) measurements and relatively high CH measurements compared to other stars in M53. 
These stars have radial velocities indicating that they are members. They also all fall along the RGB on the CMD for M53 (see Figure \ref{fig::cmd-m53}) except for one star (ID = 235), which is slightly redder than the RGB for its brightness. Because of their anomalous indices, we made additional checks to guarantee their membership. Three of the stars have more accurate radial velocities through high resolution spectroscopy from \citet{bobergm53}, and were also confirmed as members based on these radial velocity measurements. We also checked the proper motions of each star as measured by \textit{Gaia} and found that all five stars have proper motions consistent with the cluster average. One last piece of evidence we have that these stars are true members is the strengths of their Ca H and K bands, which are consistent with other stars in the cluster. 

Taking all of this evidence into account, it appears that the index measurements are accurately reflecting members with very strong CN and CH molecular bands. We offer an explanation of these stars below. To aid in our discussion, we have collected parameters of the five stars into a single table, Table \ref{tab:CH-stars}, which includes magnitudes, stellar parameters, and various abundances such as [C/Fe] and [N/Fe] from this work, and [Na/Fe], [O/Fe], [Ca/Fe], [Ti/Fe], [Ni/Fe], and [Ba/Fe] from \citet{bobergm53}. We also list the $\delta$[C/Fe] and $\delta$[N/Fe] abundances for each star, which are approximations of the initial [C/Fe] and [N/Fe] abundances of the stars before evolutionary changes due to secondary mixing along the RGB altered them (Section \ref{evolution-m53} gives more detail about these changes and how these values are calculated). 
These corrected C and N abundances allow us to compare stars in similar stages of evolution, an important factor for stars that range significantly in magnitude along the giant branch, and where C-depletion of up to 1.0 dex can occur. Table \ref{tab:CH-stars} also lists the cluster median among the RGB stars in M53 to give an easy comparison point.

\begin{figure}
\centering
\includegraphics[trim = 0.4cm 0.4cm 0.4cm 0.4cm, scale=0.43, clip=True]{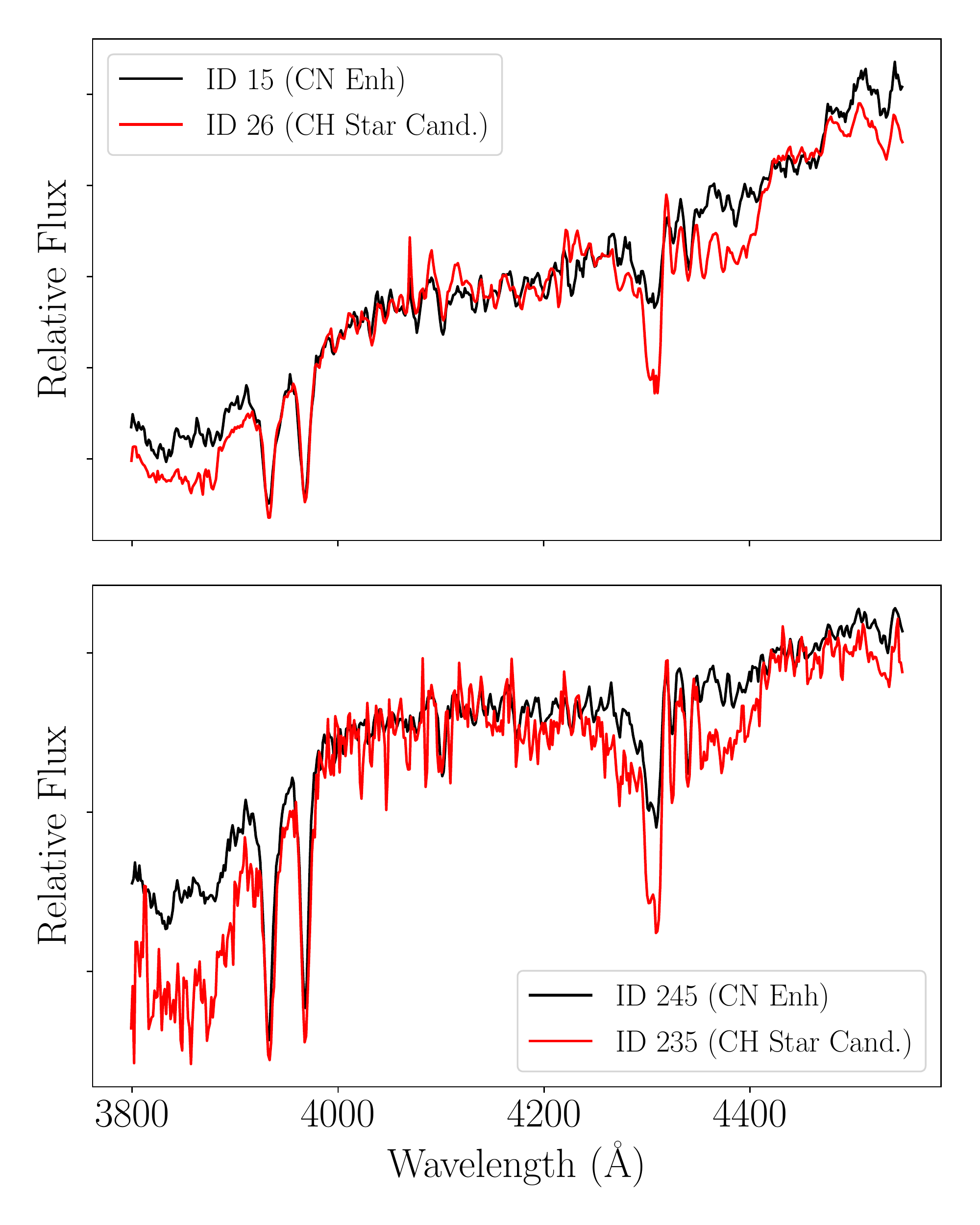}
\caption{The spectra for the two stars in our sample with the highest S(3839) measurements are plotted in red (ID = 26 and ID = 235, top and bottom panel, respectively). For comparison, stars with similar effective temperatures are also plotted in black (ID = 15 and ID = 245).}
\label{cn-extra}
\end{figure}

The spectra of the two stars with the highest S(3839) and CH index measurements, ID = 26 and 235, are shown in Figure \ref{cn-extra} where we also plot typical CN-enhanced stars of a similar temperature for comparison. The CN and CH molecular features at 3883~\AA~and 4300~\AA~are much stronger than the typical CN-enhanced star of a similar temperature. These spectra show the presence of a strong P-branch of the CH band (as indicated in the last column of Table \ref{tab:CH-stars}), which is visible immediately redward of the CH band at 4300~\AA. They also show the appearance of C$_2$ bands (Swan bands) between 4500~\AA~and 5600~\AA. 
Based on the presence of these features and the high C-enhancement, these two stars are strong candidates to be CH stars \citep[see][for an example of a classification system of C stars]{keenan93}. 
We also note, as seen in the summary of properties in Table \ref{tab:CH-stars}, that these two stars have the largest enhancements in [C/Fe] among the possible CH stars.

The other three stars do not show C$_2$ features as seen in ID = 26 and 235, but the P-branch of the CH band is still present in stars ID = 18 and 149, although it is weaker than those shown in Figure \ref{cn-extra}. In addition, these stars are both relatively enhanced in C as indicated by their $\delta$[C/Fe] values, and ID = 18 is also slightly enhanced in the s-process element, Ba compared to the cluster median. The presence of the P-branch and lack of C$_2$ features in these stars is similar to what is found in ``weak CH stars" that have been identified in clusters such as M14 and NGC 6242 \citep{cote97,sharina12}. \citet{sharina12} suggests that these CH stars with more moderate enhancement may be those that go through a different formation scenario where the enhancements are a result of incomplete CN processing as mentioned above. The final star of the five in the sample with $\delta$CN $>$ 0.2 is ID = 47. This star is the weakest candidate to be a CH star as it shows no evidence of C$_2$ bands or the P-branch, and is not strongly enhanced in C or Ba. However, its CH measurement is still on the high side of the distribution for its given T$_{eff}$, so we have left it in the sample of five possible CH stars due to its relatively high CN and CH band measurements.

Further studies of the s-process elements of these stars would be able to test and confirm whether or not they are true CH stars that have gone through some sort of interaction and mass exchange. If these stars are confirmed as CH stars, some explanation would be needed for how so many of them were found in one cluster. It could be a result of our large sample size in a cluster with the parameters necessary to allow surviving binaries to evolve into CH stars as described above. 
However, such speculation should be left until these candidate CH stars are confirmed with future high resolution spectroscopy observations. 

 \begin{deluxetable*}{cccccccccccccccccc}
 \rotate
\tabletypesize{\scriptsize}
\tablecolumns{3}
\tablewidth{0pt}
\tablecaption{\\CH Star Candidates in M53
\label{tab:CH-stars}}
\tablehead{\\
\colhead{ID$^{1}$} & \colhead{$V^{1}$} & \colhead{($B-V$)$^{1}$} & \colhead{R/R$_h$} & \colhead{M$_\mathrm{V}^{2}$} & \colhead{$T_{eff}$} & \colhead{logg} & \colhead{[C/Fe]} & \colhead{$\delta$[C/Fe]} & \colhead{[N/Fe]} & \colhead{$\delta$[N/Fe]} & \colhead{[Na/Fe]$^{3}$} & \colhead{[O/Fe]$^{3}$} & \colhead{[Ca/Fe]$^{3}$} & \colhead{[Ti/Fe]$^{3}$} & \colhead{[Ni/Fe]$^{3}$} & \colhead{[Ba/Fe]$^{3}$} & \colhead{P-Branch$^{4}$}}
\startdata
18 & 14.01 & 1.32 & 0.998 & -2.31 & 4162 &  0.86 & -0.31 & 0.27 &    1.55 & 1.44 &    0.2 &   0.4 & 0.32 & 0.18 & 0.04 & 0.40 & weak\\
26 & 14.08 & 1.32 & 0.921 & -2.24 & 4162 & 0.89 & 0.13 & 0.69 &    1.18 & 1.07 &    \nodata & \nodata & \nodata & \nodata & \nodata & \nodata & yes\\
47 & 14.52 & 1.07 & 0.796 & -1.80 & 4385 & 1.22 & -0.43 & 0.03 &    1.51 & 1.42 &    0.3 &   0.1 & 0.39 & 0.31 & 0.05 & 0.20 & no\\
149 & 15.58 & 0.92 & 2.638 & -0.74 & 4652 & 1.80 & -0.06 & 0.16 &    1.59 & 1.55 &    \nodata & \nodata & \nodata & \nodata & \nodata & \nodata & weak\\
235 & 15.96 & 0.92 & 2.165 & -0.36 & 4652 & 1.95 & 0.33 & 0.45 &    1.44 & 1.42 &    \nodata & \nodata & \nodata & \nodata & \nodata & \nodata & yes\\
\hline
Cluster Med. & & & & & & & -0.18 & 0.05 & 1.27 & 1.23 & 0.0 & 0.4 & 0.36 & 0.20 & -0.05 & 0.10\\
\enddata
\tablecomments{1. \citet{rey98}, 2. Assumes (m -- M)$_{v}$ = 16.32 \citealt{harris} (2010 edition), 3. \citet{bobergm53}, 4. Indicates the presence of the P-Branch of the CH band.}
\end{deluxetable*}

\subsection{Spatial Distribution of Multiple Populations}\label{spatial-m53}


Since our sample in M53 covers a broad range in distances from the cluster center, from 0.576 to 6.46 half-light radii (using R$_{h}$ = 1.31 arcmin; \citet{harris}: 2010 edition), we are able to study the spatial distributions of the two populations that we identified. 
Current formation models of multiple populations predict that second generation stars form in a more centrally concentrated distribution than first generation stars \citep[see e.g.,][]{dercole2008,decressin2007a,decressin2007b,bekki2010,calura19}. 
However, dynamical evolution causes the populations to become more spatially mixed over time. Specifically, relaxation-driven mass-loss has a large effect on whether or not the populations of a cluster will be spatially mixed. \citet{vesperini2013} found that unless a cluster had lost at least 60-70\% of its initial mass due to the effects of relaxation, it would still show evidence of the second generation being more centrally concentrated than the first. 

\begin{figure}
\centering
\includegraphics[trim = 0.2cm 0.4cm 0.4cm 0.2cm, scale=0.43, clip=True]{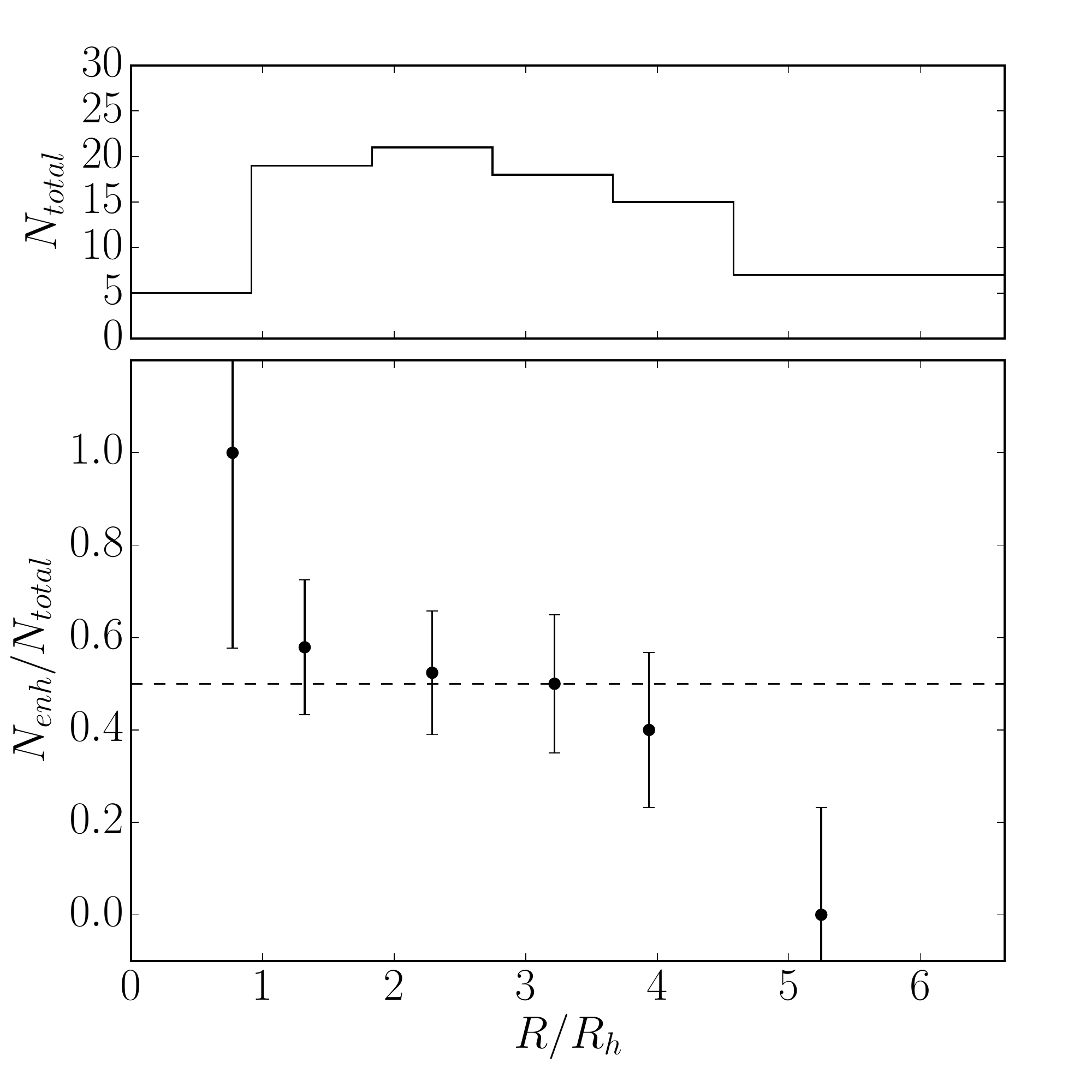}
\caption{The radial distribution of the ratio of the number of second generation RGB stars to the total number of RGB stars in M53. To classify stars, we used [N/Fe] = 1.28 dex as the dividing line between populations based on the GMM fit in Figure \ref{gmm-m53}. The top panel shows a histogram of the total number count in each bin. Bin sizes are 1.5 arcmin out to 4.5 half-light radii. All stars at a greater radius in our sample are combined into one bin. The points shown in the bottom panel are centered at the average position of the stars in a given radial bin. The error bars for these points reflect the uncertainty in the ratio based on counting statistics. A black dashed line indicates our average ratio for the entire sample.}
\label{radial-m53}
\end{figure}

M53 has a long relaxation time (log(t$_{rh}$)=9.76; \citealt{harris}: 2010 edition), is currently located at a large galactocentric distance (R$\sim$18.4 kpc; \citealt{harris}: 2010 edition), and estimates of its orbital parameters based on \textit{Gaia} data suggest it never moves close to the central regions of the Galaxy (R$_{peri}\sim$9 kpc, R$_{apo}\sim$22 kpc; \citealt{baumgardt19}). Finally, a recent study by \citet{ebrahimi20} has found a steep slope for the stellar mass function in M53 suggesting that the cluster has not been significantly affected by mass loss due to two-body relaxation. These parameters indicate that M53 is unlikely to have suffered a strong internal evolution and mass loss due to two-body relaxation and might thus have preserved some memory of the initial differences between the spatial distribution of first and second generation stars \citep{vesperini2013}. Our analysis of the radial distribution of the ratio of the number of second generation stars to the total number of stars, shown in Figure \ref{radial-m53}, does indeed suggest that second generation stars are more centrally concentrated than first generation stars although a larger number of stars and a broader radial coverage would be needed to confirm the statistical significance of the trend we found.

\citet{bobergm53} reached a similar conclusion from the cumulative distribution based on their results separating populations based on Na abundance. 
Dividing their sample into an inner and outer bin separated at R$=$3 R$_h$, we find that the percentage of Na-enhanced stars is 28.4\% in the inner bin and 4.8\% in the outer bin, indicating a more centrally concentrated second generation. We discuss the differences between the percentage of second generation stars found by our $\delta$CN identification method and \citet{bobergm53} in the following section. 


\subsection{Comparison to Other Methods of Identifying Multiple Populations}

\subsubsection{Comparison with Na-O Anti-Correlation} 


As mentioned in \ref{targ-select}, our sample included stars that had observations of Na and O from \citet{bobergm53}, so that we could compare our identification of multiple populations with the identification based on the Na-O anti-correlation. Comparisons between these two classification systems identify outliers, which help constrain theories on multiple population formation. This comparison is also interesting in M53 because \citet{bobergm53} found three stars in their sample to be Na-normal, but CN-enhanced based on $\delta$CN values from \citet{martell}. We wanted to explore if these outliers continued to exist with updated measurements of $\delta$CN and [N/Fe], and if additional Na-normal, CN-enhanced stars would be found in a larger sample of stars.

\begin{figure}
\includegraphics[trim = 0.2cm 0.4cm 0.4cm 0.2cm, scale=0.33, clip=True]{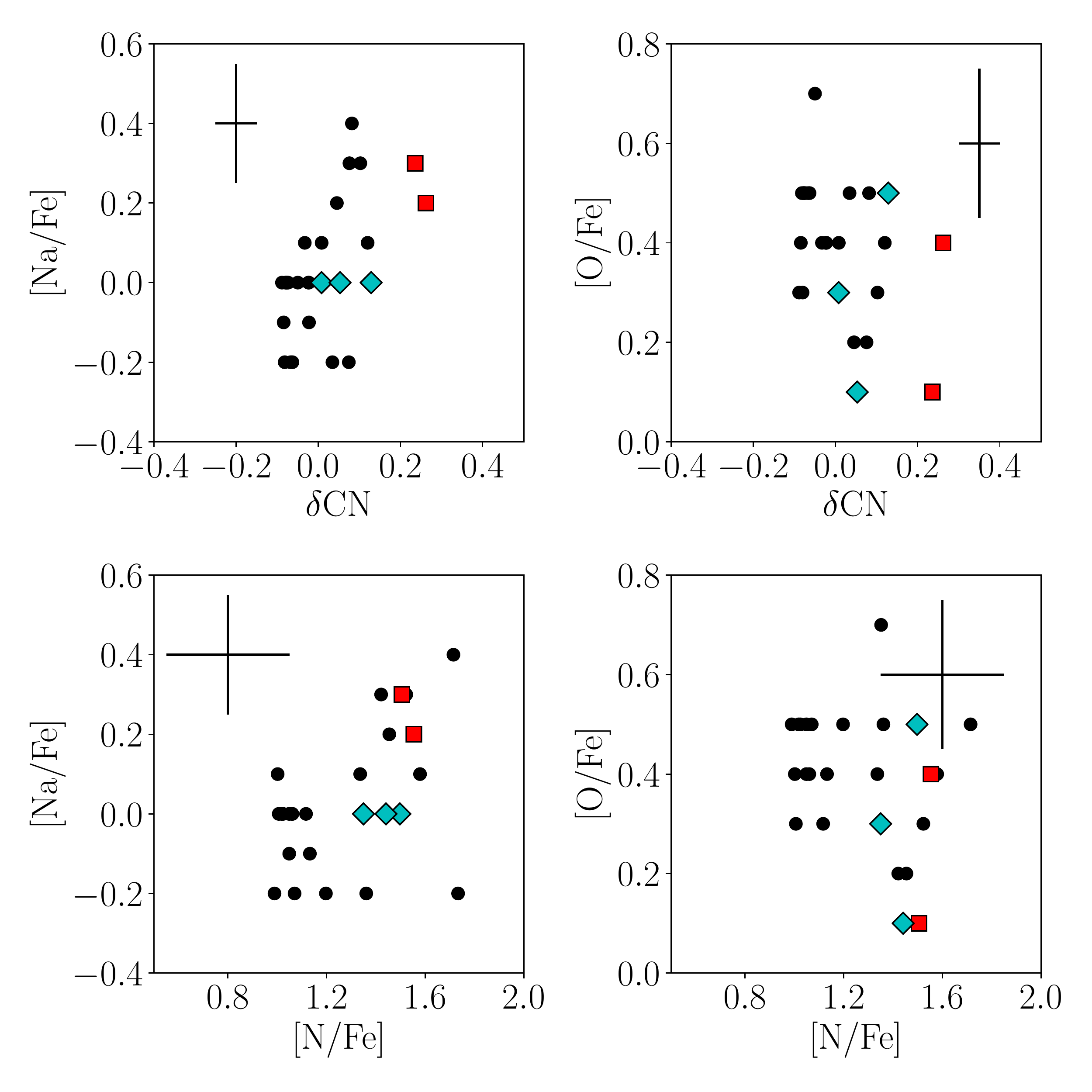}
\caption{\textbf{Top:} [Na/Fe] and [O/Fe] vs $\delta$CN (left and right, respectively). Na and O measurements are from \citet{bobergm53}. \textbf{Bottom:} Same as above vs. [N/Fe]. RGB stars are shown as black circles, and CH star candidates are shown as red squares. Cyan diamonds indicate stars that were identified by \citet{bobergm53} as stars with possible high N and low Na values. Representative error-bars show the uncertainties with each measurement.}
\label{nafe-m53}
\end{figure}

We plot [Na/Fe] and [O/Fe] measurements from \citet{bobergm53} vs. our $\delta$CN and [N/Fe] values in Figure \ref{nafe-m53}. 
Representative uncertainties are also shown in each panel. A general correlation between N-Na and an anti-correlation between N-O is observed as expected, although neither relation is very tight due to the relatively large uncertainties in such a low metallicity cluster. We note that there are some stars with intermediate [Na/Fe] values and slightly enhanced $\delta$CN or [N/Fe] values. Included among these stars are the three stars observed by \citet{bobergm53} to have high $\delta$CN and low [Na/Fe], which are indicated as blue diamonds in Figure \ref{nafe-m53}. In addition, we find two stars with [Na/Fe] as low as -0.2 dex that are found to have enhanced $\delta$CN and [N/Fe] values. However, none of these stars deviate from the correlation by more than one sigma, which makes it difficult to draw strong conclusions about whether or not they indicate a population of stars with enhanced N but normal Na abundance. Therefore, our results would indicate that our method using CN band analysis to identify multiple populations and the analysis using the Na-O anti-correlation are in general agreement with one another.

While the Na, O, and N values are in general agreement, \citet{bobergm53} found a different ratio between first and second generation stars than we do. Thirteen stars were identified as second generation out of their total sample of 49 stars, which gives 27\% (compared to our 50\%). Other studies have also found that the ratio between populations in clusters does not match what is found when stars are sorted using the Na-O anti-correlation. For example in a study of 12 GCs, \citet{pancino2010} found that all clusters studied had a higher percentage of first generation stars compared to what was observed using the Na-O anti-correlation. \citet{smith2013} and \citet{smith2015} also observed stars in M5 and 47 Tuc that were Na normal, but CN enhanced, and vice versa (in the case of M5). Anomalous stars such as these could explain the differences in percentage of first generation stars based on identification method seen in \citet{pancino2010}. 

However, in the case of M53, the discrepancy between the ratio between first and second generation stars likely comes from the fact that the [Na/Fe] distribution, as seen in Figure 6 of \citet{bobergm53}, is more continuous than the [N/Fe] distribution. This means that the two populations do not separate clearly in the Na-O plane as they do in the [N/Fe] distribution. We also note that moving the separating line in [Na/Fe] from 0.2 dex, which was used by \citet{bobergm53} to distinguish populations, to 0.0 dex would bring the ratio in line with our 50\%. In addition, the average uncertainties of the [Na/Fe] measurements are 0.15 dex, and lead to greater uncertainties in classification. 
These factors, along with the smaller sample size used by \citet{bobergm53}, could explain how different ratios would be found.


\subsubsection{N-Na Correlation}
\begin{figure}
\centering
\includegraphics[trim = 0.2cm 0.2cm 0.2cm 0.2cm, scale=0.45, clip=True]{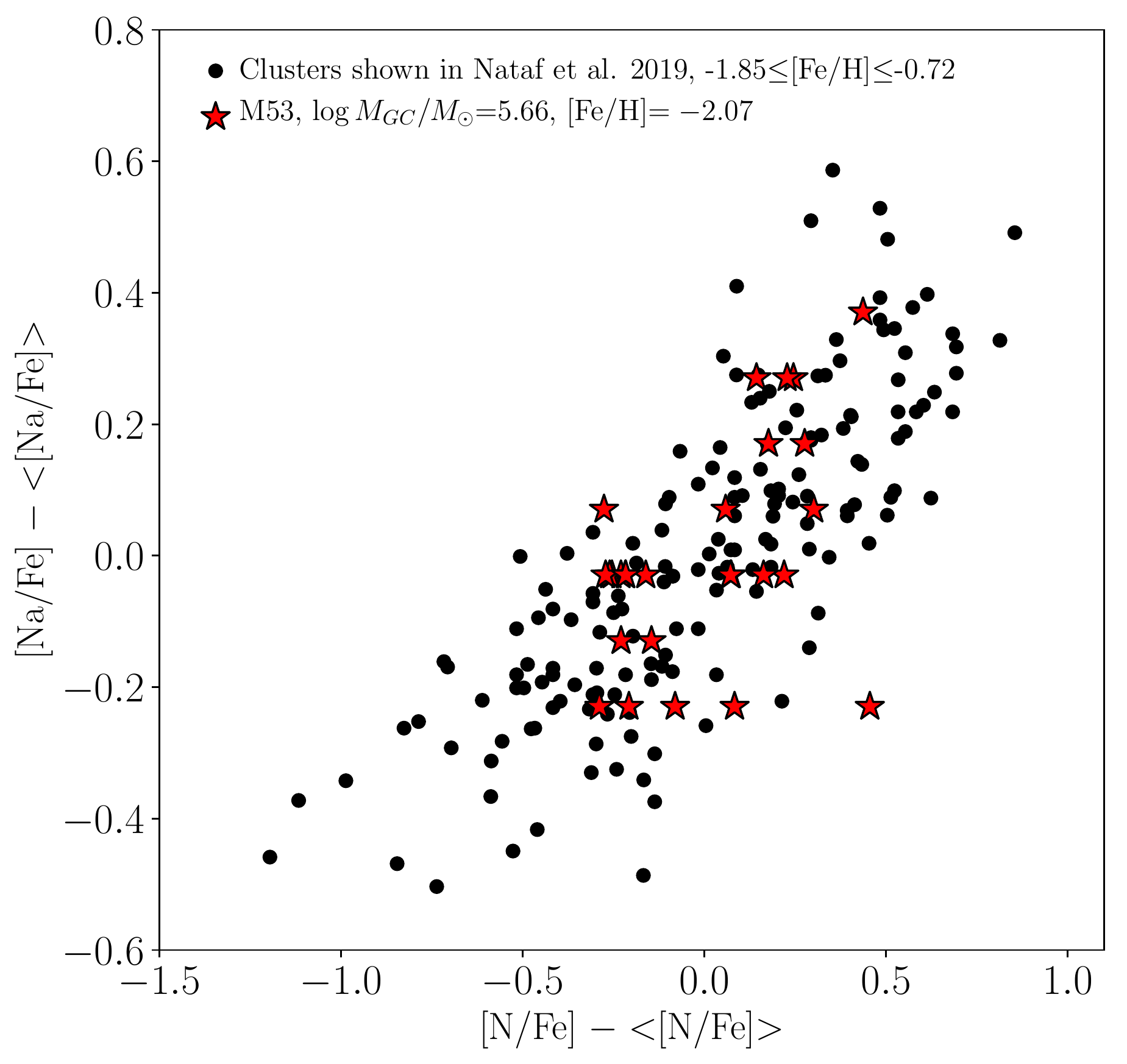}
\caption{Figure 19 from \citet{nataf19} with our data for M53 plotted as red stars. The x-axis shows the difference between [N/Fe] for each star and the average [N/Fe] for the cluster, and the y-axis shows the same for Na. M53 shows a similar correlation as the other clusters, which agrees with the results of the analysis of \citet{nataf19} showing that this correlation is independent of cluster mass.}
\label{navsn}
\end{figure}

A recent study by \citet{nataf19} compared correlations in N, Na, and Al in a large sample of clusters to determine if the strength and slope of these correlations were related to cluster properties such as mass and metallicity. As part of this study, they compared the slope in the correlation between [Na/Fe] and [N/Fe] among clusters of varying total mass and metallicity to test if the slope was related to either of these parameters. They found that the slope does not appear to change from cluster to cluster, which would indicate that it is independent of cluster mass and metallicity. We add our sample in M53 to this study as shown in Figure \ref{navsn} where we plot [Na/Fe] - $<$[Na/Fe]$>$ vs. [N/Fe] - $<$[N/Fe]$>$ for the stars in M53 (this study) with the values for other clusters from \citet{nataf19} (see references therein). [x/Fe] - $<$[x/Fe]$>$ is simply the value for N or Na for a star subtracted by the average value for the cluster. As in \citet{nataf19}, these values are calculated to compensate for the difference in [Fe/H] between the clusters, and to put all of the measurements on approximately the same scale for better comparison. Figure \ref{navsn} shows that the slope in the correlation between Na and N for M53 generally agrees with those observed for other clusters considering the uncertainties in our measurements and those in [Na/Fe]. Our results for M53 then support the conclusion of \citet{nataf19} that the slope in the correlation between N and Na appears to be independent of cluster mass and metallicity.

\subsubsection{Comparison with HST UV Legacy Archive Photometry}


HST UV photometry has been used to identify multiple populations in M53 by \citet{miloneatlas}. This method makes use of pseudo-colors, or different combinations of colors, that enhance the separation between the populations in a cluster because they are dependent on C, N, and O abundances of a star \citep[see][]{piotto,miloneatlas}. A ``pseudo-CMD" created using one of these pseudo-colors shows distinct RGB sequences, where the second generation stars rich in N appear ``bluer" and the first generation stars, with lower N, appear ``redder."

\begin{figure}
\centering
\includegraphics[trim = 0.4cm 0.4cm 0.4cm 0.4cm, scale=0.42, clip=True]{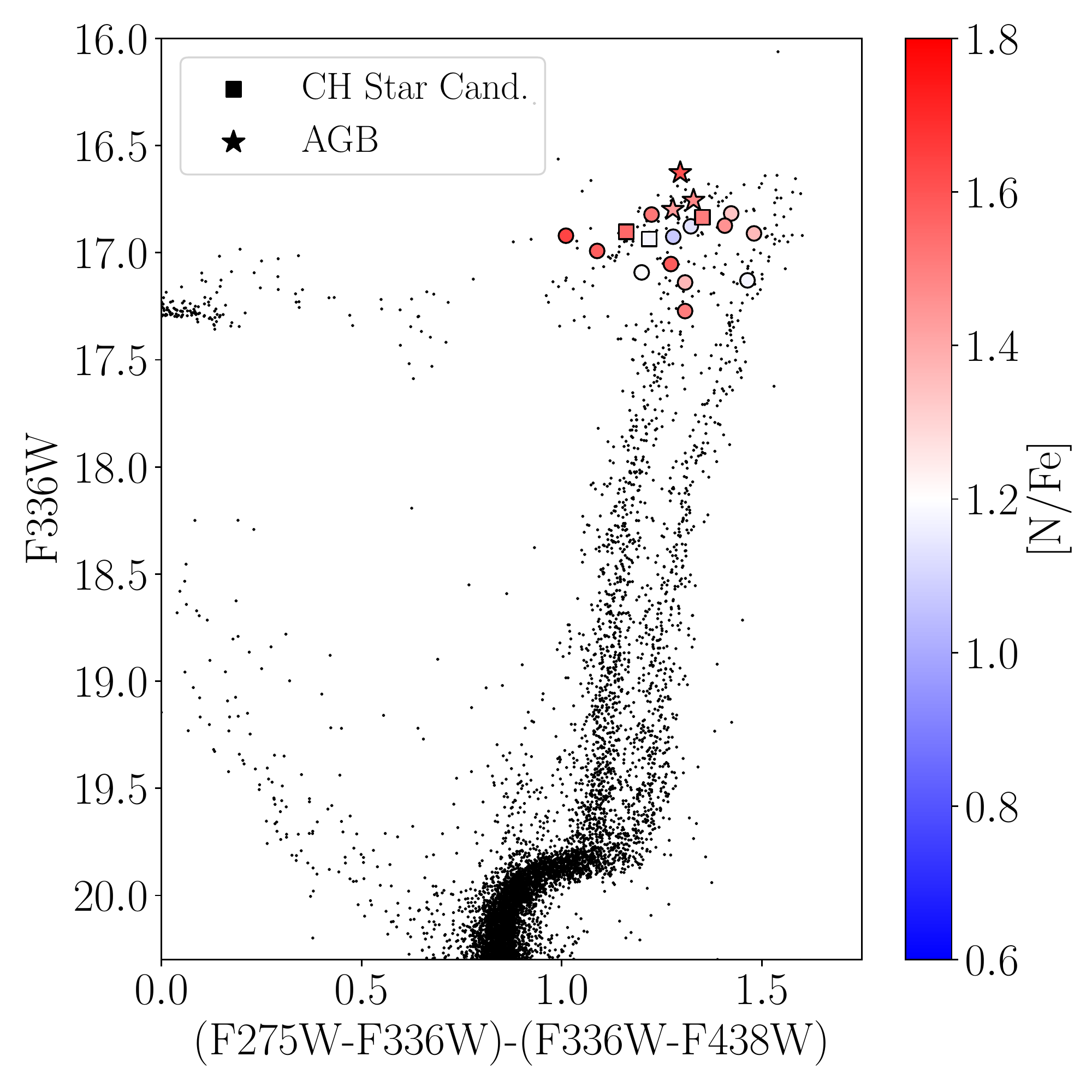}
\caption{A pseudo-color magnitude diagram \citep{piotto,nardiello2018} to determine multiple populations photometrically shown with stars color-coded based on [N/Fe]. RGB stars are shown as circles, AGB stars as stars, and candidate CH stars as squares. The pseudo-color enhances the separation between the red giant branches of the populations in the cluster, but the sequences become confused at the brightest magnitudes.}
\label{hst-m53}
\end{figure}

Using HST photometry from \citet{piotto} and \citet{nardiello2018}, we re-create the pseudo-CMD in M53 to make a comparison between our method of identifying multiple populations and the method used by \citet{miloneatlas}. We have only 19 stars in common with the data from the HST photometry because the HST photometry covers a much narrower field than our data. Figure \ref{hst-m53} shows these 19 stars color-coded by [N/Fe] strength overplotted on the pseudo-CMD. The brighter, cooler stars along the RGB in the pseudo-CMD move to the ``blue" and the RGB bends over as their magnitudes in these HST UV filters become very faint. Essentially, these cool, red stars are not bright enough in the bluer passbands of the near UV filters to allow for an accurate enough measurement of their pseudo-color to confidently sort them into populations. Unfortunately, because the HST photometry goes much deeper than our sample, the stars in common are some of the brightest in the HST photometry. Therefore, most of our stars with HST photometry fall in a location where the two RGB sequences are no longer distinguished. The faintest four stars in common with the HST photometry appear to fall on the correct RGB sequences, but we are unable to make a strong conclusion about the other RGB stars in our sample with HST photometry. 
We also do not have enough AGB stars in common with the HST photometry to make any strong conclusions about their locations.

\citet{miloneatlas} finds the cluster to be 67$\pm$2\% second generation stars over their radial distance of 1.35 half-light radii. From Figure \ref{radial-m53}, we find that the ratio of CN-enhanced stars is around 70\% out to this radial distance, consistent with the result from \citet{miloneatlas} within the same radius. In Section \ref{spatial-m53}, we showed that the second generation is more centrally concentrated in M53, which explains why the percentage of second generation stars found by \citet{miloneatlas} differs from the global value calculated here using the total sample covering a wider radial range.

\subsection{Effect of Evolution on C and N Abundances} \label{evolution-m53}


RGB stars in GCs brighter than the LFB have been shown to have decreasing surface C abundances and increasing surface N abundances as a function of magnitude. Previous studies have tried to constrain the rate of C depletion by studying the brightest RGB stars in GCs and determining a rate based on an assumed initial C abundance \citep{martellmixing,gerber19}. However, because these stars can be observed to have a large range of initial C abundances, these studies have struggled to calculate precise C-depletion rates. 

\begin{figure*}
\centering
\includegraphics[trim = 0.4cm 0.4cm 0.4cm 0.4cm, scale=0.4, clip=True]{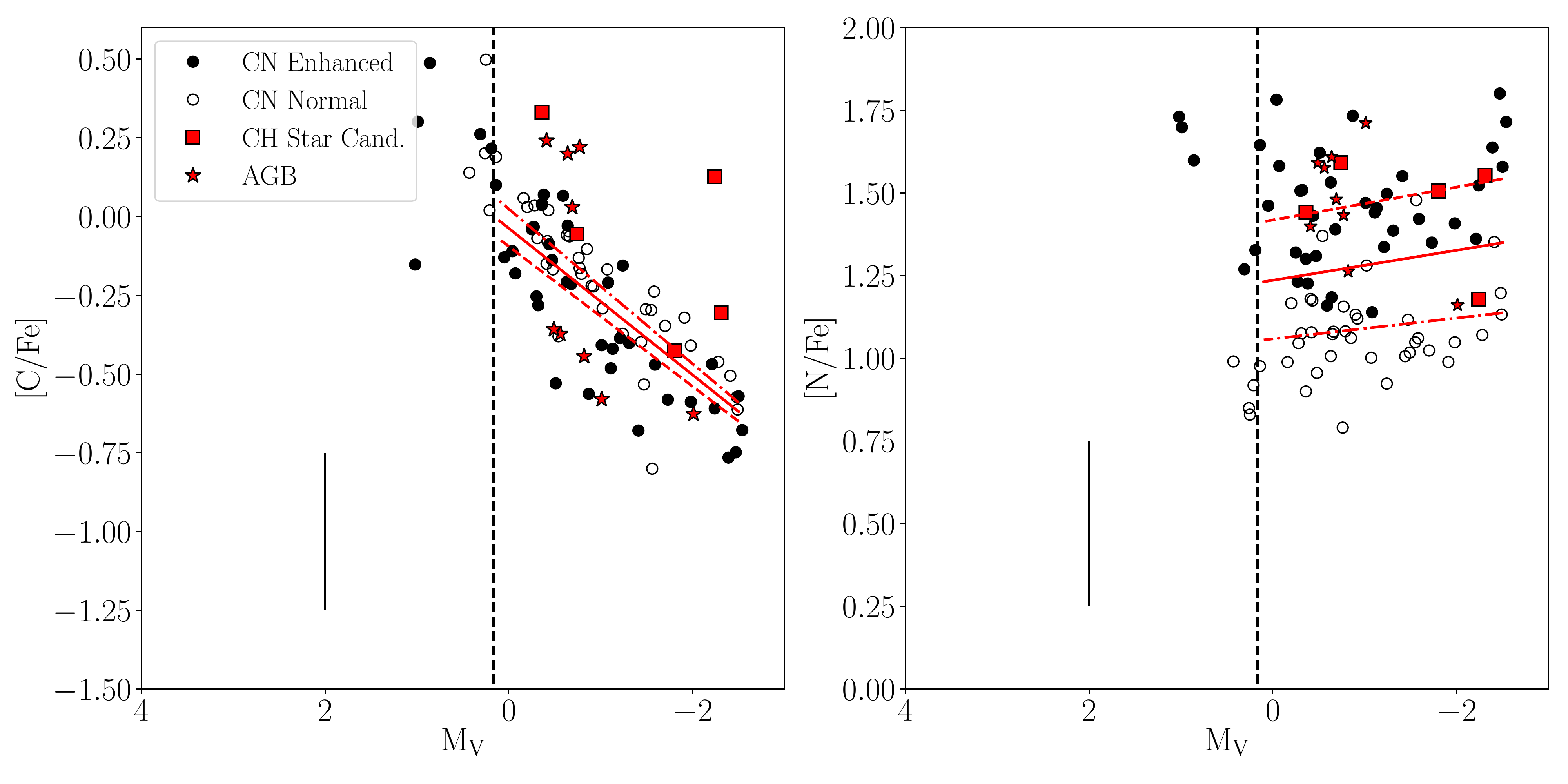}
\caption{[C/Fe] and [N/Fe] vs. M$_\mathrm{V}$ for all 94 member stars in M53. Stars are coded based on their $\delta$CN with CN-enhanced stars as filled circles and CN-normal stars as open circles. AGB stars are shown as red stars, and CH star candidates as red squares. A dashed line at $M_\mathrm{V} = 0.0168$ represents the LFB \citep{lfb}, where secondary mixing is expected to begin for this cluster. Linear fits to RGB stars are shown in red for CN-normal, CN-enhanced, and both populations as dot-dashed, dashed, and solid lines, respectively. Representative error bars are shown in the bottom right of each plot.}
\label{xfevsm-m53}
\end{figure*}

We avoid these issues and can determine a more precise C-depletion rate in M53 due to our large sample size covering a range in magnitudes from the LFB to the tip of the RGB. Our sample allows for each population in M53 to be analyzed independently to make comparisons. To determine a rate of mixing for M53, we use a simple linear-least squares regression for our entire sample that begins at the LF bump measured to be M$_{bump}$ $\sim$~0.02 by \citet{lfb}. We also fit the change of [C/Fe] and [N/Fe] as a function of magnitude for the CN-enhanced and CN-normal populations independently. We do not include the AGB stars or candidate CH stars in our sample in any of these fits.

Figure \ref{xfevsm-m53} shows the results of these fits over-plotted on [C/Fe] and [N/Fe] vs. M$_\mathrm{V}$ for our sample. CN-enhanced stars are indicated as filled circles and CN-normal stars as open circles. Because M53 is a relatively distant and faint cluster, we were not able to observe many stars before the LFB to efficiently constrain the initial C and N abundances of stars before mixing occurs, which means we cannot comment on the whether or not there is pre-LFB bump mixing as reported by \citet{angelou2012}. However, we are still able to determine the rate of depletion because we did observe stars faint enough to reach the LFB and capture the range of magnitudes over which stars are mixing in M53.

Our analysis shows that both populations in M53 have similar C-depletion and N-enhancement rates, which indicates similar mixing efficiencies. For the CN-enhanced population, we find a C-depletion rate of 0.22 $\pm$ 0.03 dex mag$^{-1}$, and for the CN-normal population, we find a depletion rate of 0.25 $\pm$ 0.03 dex mag$^{-1}$. For [N/Fe], we find enhancement rates of -0.05 $\pm$ 0.03 dex mag$^{-1}$ and -0.03 $\pm$ 0.04 for the CN-enhanced and normal populations, respectively. We list these rates in Table \ref{tab:rates-m53}.

Since the rates for the change in [C/Fe] and the change in [N/Fe] for each of the two populations are consistent within the uncertainties, we conclude that both populations are mixing at the same rate. This result agrees with our earlier results for M10 where both populations also had depletion rates that agreed within uncertainties. While the rates of C-depletion do not match the inverse rate of N-enhancement, this discrepancy can be explained by the logarithmic nature of the [N/Fe] and [C/Fe] measurements and the fact that the stars have much higher levels of N than C. For example, consider a parameter whose value changes by 3. If the difference is from 1 to 4, then on a log scale the parameter ranges from 0 to 0.6 dex, but if the difference is from 10 to 13, on a log scale it is only 0.11 dex.


 \begin{deluxetable}{ccc}
\tabletypesize{\footnotesize}
\tablecolumns{3}
\tablewidth{0pt}
\tablecaption{\\Rate of Abundance Change vs. Magnitude for M53 and M92
\label{tab:rates-m53}}
\tablehead{
\colhead{\textbf{Sample}} & \colhead{\textbf{$d$[C/Fe]/$dM_\mathrm{V}$}} & \colhead{\textbf{$d$[N/Fe]/$dM_\mathrm{V}$}}}
\startdata
\textbf{M53} & &\\
\hline
All & 0.23 $\pm$ 0.02 & -0.05 $\pm$ 0.04\\ 
CN-enhanced & 0.22 $\pm$ 0.03 & -0.05 $\pm$ 0.03\\
CN-normal & 0.25 $\pm$ 0.03 & -0.03 $\pm$ 0.04\\
\hline
\textbf{M92$^{1}$} & &\\
\hline
All & 0.253 $\pm$ 0.023 & \nodata \\
\enddata
\tablecomments{1. \citet{m3cfe}}
\end{deluxetable}

\begin{figure}
\centering
\includegraphics[trim = 0.4cm 0.4cm 0.4cm 0.4cm, scale=0.34, clip=True]{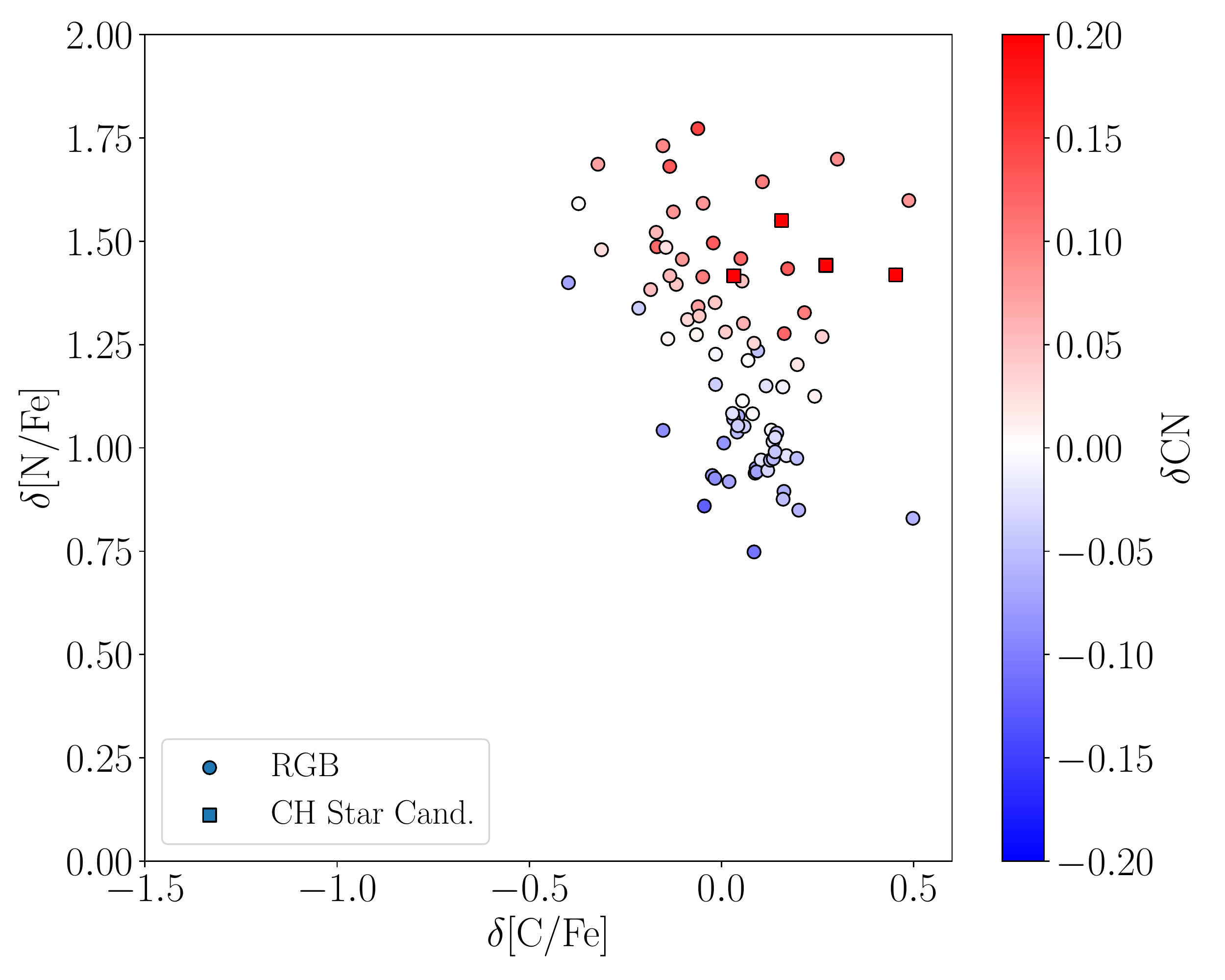}
\caption{$\delta$[N/Fe] vs. $\delta$[C/Fe] for RGB stars in our sample created by removing the trends in abundance found with magnitude, which are shown in Figure \ref{xfevsm-m53}. The color convention is the same as that used in Figure \ref{nfevscfe-m53}.}
\label{deltanfevsdeltacfe-m53}
\end{figure}

Once we calculate the rates of change for C and N, we apply these rates to our abundances to estimate what a star's initial abundance would have been before experiencing extra mixing. Normalized $\delta$[C/Fe] and $\delta$[N/Fe] measures for each star are created by using the linear fits to [C/Fe] and [N/Fe] versus magnitude to correct for these evolutionary effects. These corrected abundances are shown in Figure \ref{deltanfevsdeltacfe-m53}, and can be used to study the C-N anti-correlation in RGB stars before the evolutionary effects act on abundances. In Figure \ref{deltanfevsdeltacfe-m53}, the two populations now show a clearer separation in the C-N plane compared to Figure \ref{nfevscfe-m53}. It also seems that the second generation shows a larger spread in initial C and N abundances than the first generation.

In addition, we compare our data to similar data sets of C abundances for M92 as compiled by \citet{m3cfe} from various literature sources, and for three low metallicity GCs studied by \citet{kirby15} (NGC 2419, M68, and M15). We chose these clusters because they have comparably sized data sets in the literature, and because of their similar age and metallicity to M53. Because these clusters have similar metallicities, most theories of extra mixing predict they should have similar mixing efficiencies.

\citet{m3cfe} found a value of 0.25 $\pm$ 0.02 dex mag$^{-1}$ overall for the [C/Fe] measurements in M92, which agrees well with what we find for M53 (0.23 $\pm$ 0.02 dex mag$^{-1}$). \citet{kirby15} fit NGC 2419, M68, and M15 simultaneously since all were found to have similar rates of C depletion, and found a depletion value of $d$[C/Fe]/$d\mathrm{log(}L/L_{\odot}\mathrm{)}$ = -0.82 $\pm$ 0.02. On the same $\mathrm{log(}L/L_{\odot}\mathrm{)}$ scale, our C depletion rate for the entire M53 sample of RGB stars is $d$[C/Fe]/$d\mathrm{log(}L/L_{\odot}\mathrm{)}$ = -0.50 $\pm$ 0.05. While the rate of depletion for M53 is lower than what was found in clusters of similar metallicity by \citet{kirby15}, we note that \citet{kirby15} did not separate stars by population in two of the three clusters that they studied (M68 and M15), which make up a bulk of their sample. If their sample in either of these clusters happened to include more second generation stars than first generation at the brighter luminosity end or more first generation stars than second generation stars at the lower luminosity end (or both), then the slope of the fit to C depletion rate could end up skewed towards a more negative value since second generation stars have lower initial C abundances than first generation stars. This effect could be enough to explain the difference in depletion rates between the two fits to samples of RGB stars from GCs with similar metallicities.

\subsubsection{C+N+O Abundances}

\begin{figure*}
\centering
\includegraphics[trim = 0.0cm 5cm 0.0cm 5cm, scale=0.45, clip=True]{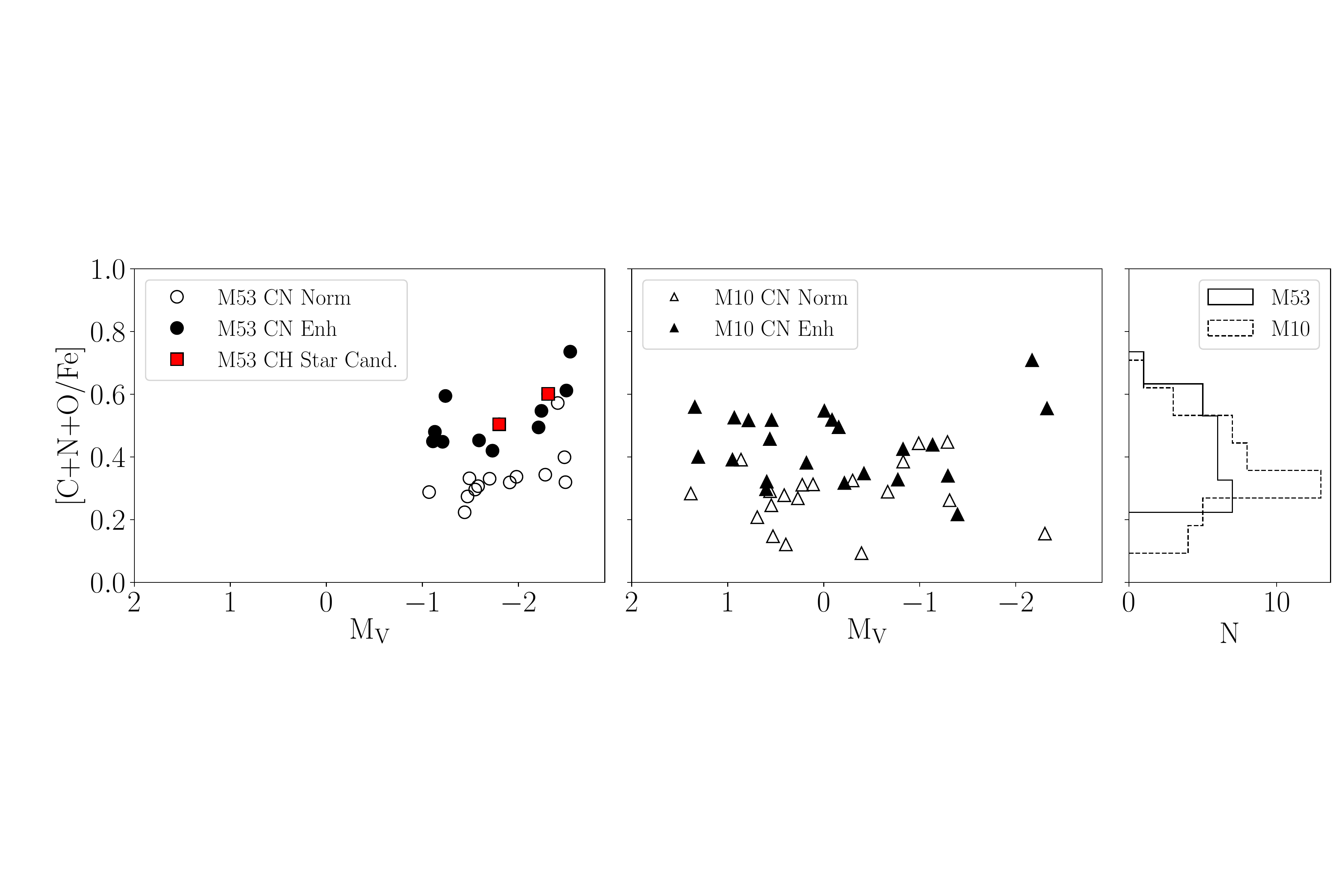}
\caption{\textbf{Left:} [C+N+O/Fe] measurements for RGB stars in our sample in M53 with O abundances from \citet{bobergm53}. The color convention is the same as that used in Figure \ref{bands-m53}. \textbf{Middle:} [C+N+O/Fe] measurements for RGB stars in M10 using data from G18 and O abundances from \citet{gir,uves}. \textbf{Right:} Histograms showing the distributions of both samples. M53 is shown as a solid line, and M10 as a dashed line.}
\label{cnofevsmv-m53}
\end{figure*}

We study the C+N+O abundance of each population in M53 by combining our C and N abundances with the O abundance from \citet{bobergm53}, and compare it with similar measurements in M10 using the values from G18. The left panel of Figure \ref{cnofevsmv-m53} shows the calculated [C+N+O/Fe] vs. M$_\mathrm{V}$ for all of the stars in our sample with O abundances. We find that the second generation has a higher C+N+O abundance than the first generation with a separation of roughly 0.2 dex \citep[see][for models predicting possible C+N+O differences similar to this value]{dantona2016}. The second generation has an average [C+N+O/Fe] value of 0.53 $\pm$ 0.09 (std) dex, and the first generation has an average of 0.33 $\pm$ 0.08 (std) dex. A student test on the hypothesis that the [C+N+O/Fe] samples for both populations were pulled from a distribution with the same mean resulted in a p value of 1x10$^{-5}$, which indicates that the two populations have significantly different [C+N+0/Fe] abundances. 

These results for the [C+N+O/Fe] in each population in M53 are comparable to what we observed in M10 using our C and N measurements and the O abundances from the literature (G18 and references therein). In the middle panel of Figure \ref{cnofevsmv-m53}, we plot [C+N+O/Fe] for M10 as a function of magnitude. For M10, we find the second generation has an average value of 0.44 $\pm$ 0.11 (std) dex, and the first generation has an average value of 0.28 $\pm$ 0.10 (std) dex. We conducted the same student test on the [C+N+O/Fe] values for M10 as we did for M53, which resulted in a p value of 3x10$^{-5}$. This p value indicates that the two populations in M10 also have significantly different [C+N+0/Fe] abundances. The average [C+N+O/Fe] values for each population in M10 and M53 are shown in Table \ref{cnotable}. The right panel of Figure \ref{cnofevsmv-m53} shows the distribution of C+N+O for each cluster for a direct comparison. 

 \begin{deluxetable}{ccc}
\tabletypesize{\footnotesize}
\tablecolumns{3}
\tablewidth{0pt}
\tablecaption{[C+N+O/Fe] and [C+N/Fe] for Populations in M53 and M10
\label{cnotable}}
\tablehead{
\colhead{\textbf{Population}} & \colhead{\textbf{[C+N+O/Fe]}} & \colhead{\textbf{[C+N/Fe]}}}
\startdata
\textbf{M53} & &\\
\hline
CN-enhanced & 0.53 $\pm$ 0.09 (std) & 0.81 $\pm$ 0.15 (std)\\
CN-normal & 0.33 $\pm$ 0.08 (std) & 0.47 $\pm$ 0.10 (std)\\
\hline
\textbf{M10$^{1}$} & &\\
\hline
CN-enhanced & 0.44 $\pm$ 0.11 (std) & 0.71 $\pm$ 0.21 (std)\\
CN-normal & 0.28 $\pm$ 0.10 (std) & 0.19 $\pm$ 0.12 (std)\\
\enddata
\tablecomments{1. G18 and references therein}
\end{deluxetable}

\begin{figure*}
\centering
\includegraphics[trim = 0.0cm 5cm 0.0cm 5cm, scale=0.45, clip=True]{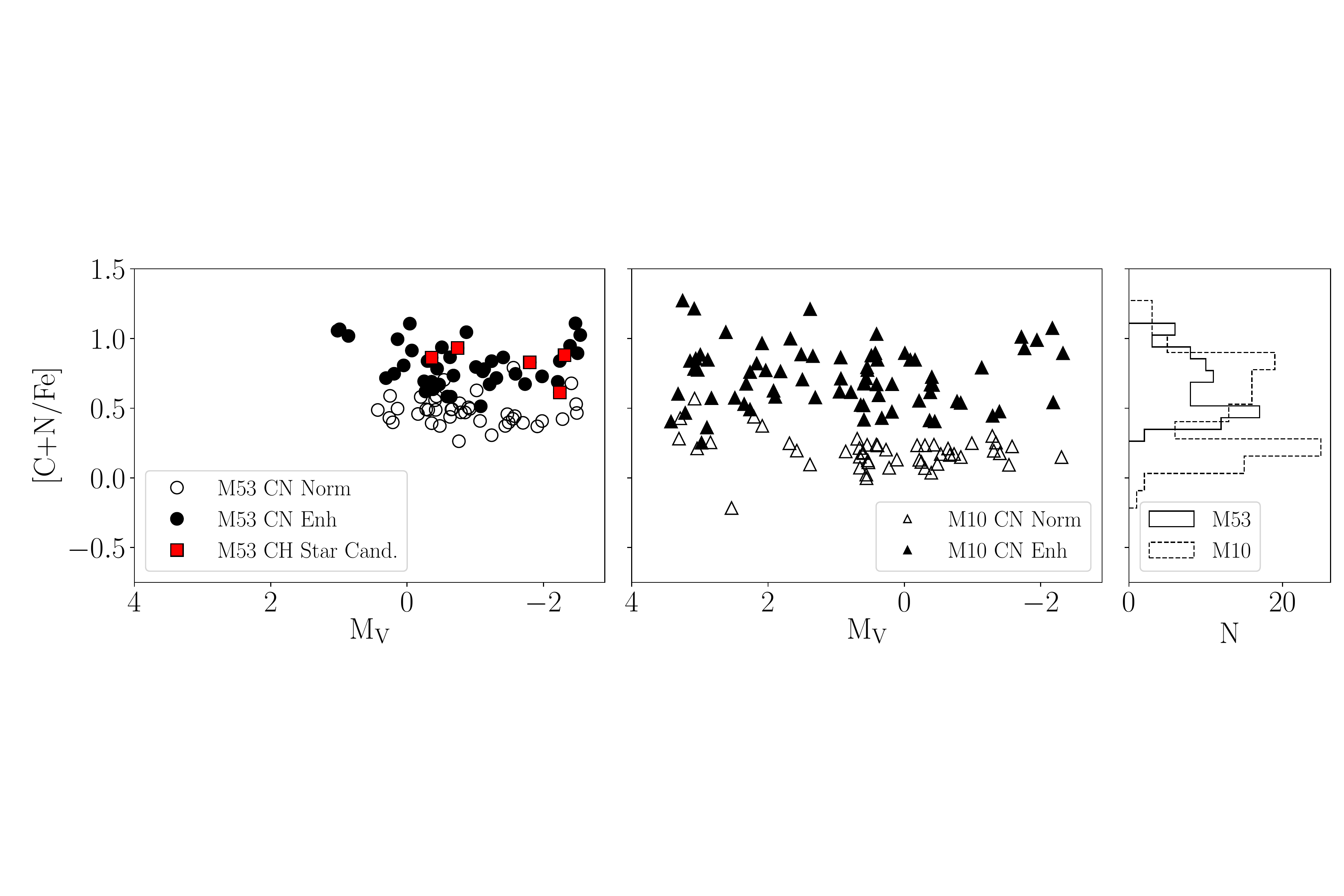}
\caption{\textbf{Left:} [C+N/Fe] measurements for RGB stars in our sample in M53. The color convention is the same as that used in Figure \ref{bands-m53}. \textbf{Middle:} [C+N/Fe] measurements for RGB stars in M10 using data from G18. \textbf{Right:} Histograms showing the distributions of both samples. M53 is shown as a solid line, and M10 as a dashed line.}
\label{cnfevsmv-m53}
\end{figure*}

The difference in [C+N+O/Fe] between populations and the standard deviation in each population in M10 is similar to what we measure in M53. Both populations in each cluster also have approximately equal standard deviations. 
We also note that the C+N+O abundances are not heavily affected by the extra mixing that is causing C-depletion and N-enhancement due to the fact that the material being brought to the surface of the star has been processed through the CN(O)-cycle. Therefore, the C being depleted has been converted into N, which keeps C+N+O constant. We study the C+N abundances to further demonstrate how the sum of these abundances stays constant with increasing luminosity. The left panel of Figure \ref{cnfevsmv-m53} shows the [C+N/Fe] values for all RGB stars in our sample in M53, and the middle panel shows the [C+N/Fe] values for M10 as in Figure \ref{cnofevsmv-m53}. From this figure, we find that there is no trend with magnitude for either cluster as expected. In each cluster, the second generation also has a slightly larger spread in [C+N/Fe] than the first generation with the populations having standard deviations of 0.15 dex and 0.1 dex, respectively in M53, and 0.2 dex and 0.1 dex in M10. The dispersions about the mean for the first generation in each cluster are consistent with our uncertainties for [C/Fe] and [N/Fe], while the dispersions for the second generation in each cluster are slightly larger. The mean abundance differences are given in Table \ref{cnotable}.

\section{Conclusions} \label{Conclusions-m53}
In this paper, we focused on measurements of CN and CH band strengths for 85 RGB stars and 9 AGB stars in the GC M53. We used these measurements to identify CN-enhanced and CN-normal populations, and calculate C and N abundances for all stars in the sample. The conclusions we reached from this analysis are listed below:
\begin{enumerate}
\item We find a large spread in CN at a given T$_{eff}$ that indicates the existence of more than one population in the cluster. The populations separate more clearly in the [N/Fe] distribution, which indicates the presence of a N-enhanced and a N-normal population. We find the RGB stars in our sample to be evenly divided between each population.
\item We have identified 5 stars as CH star candidates based on their extremely strong CN and CH bands. Of these, the strongest 2 candidates have a clear P-branch of their CH band, as well as the slight appearance of molecular C$_2$ features. We have thoroughly checked the membership of these stars with radial velocities determined through high resolution spectroscopy and proper motions from \textit{Gaia}. Stellar properties and abundances are presented for these stars in Table \ref{tab:CH-stars}. Follow up high resolution spectroscopy of these stars is necessary to confirm whether these are indeed CH stars that have gone through some sort of interaction with a binary companion. If they are confirmed to be CH stars, having so many in one cluster may require further explanation.
\item We compare our method of identifying multiple populations to other methods such as the Na-O anti-correlation \citep{bobergm53} and HST UV photometry \citep{miloneatlas}. Our method agrees reasonably well with both of these methods. While the correlation and anti-correlation between N-Na and N-O, respectively, are not as tight as they were in M10, this can be explained by the higher uncertainties involved in abundance measurements for a lower metallicity cluster.
\item We find that the second generation (CN-enhanced population) in M53 is more centrally concentrated than the first generation. This result agrees with the conclusions of \citet{bobergm53} using the Na-O anti-correlation to identify populations. Various dynamical parameters suggest that M53 is not in its advanced evolutionary stages and has not suffered strong mass loss due to relaxation, and the presence of a radial gradient in the fraction of second generation stars is generally consistent with the dynamical models of \citet{vesperini2013}.
\item A large range of values is found for both the [C/Fe] and [N/Fe] measurements for the RGB stars, which is expected. Because most of our stars have evolved past the LFB, the magnitude dependence of [C/Fe] and [N/Fe] abundances obscures the distinct populations in the C-N plane. The separation between these two populations becomes clearer once we remove the dependence on magnitude based on our fits to [C/Fe] and [N/Fe] vs. magnitude.
\item We calculate [C+N+O/Fe] for stars in our sample that have O abundances from \citet{bobergm53}, and compare to similar measurements in M10 from G18 and references therein. In M53, we find second generation stars to be enhanced in [C+N+O/Fe] by $\sim$0.2 dex, an enhancement comparable to that found for second generation stars in M10.
\item Our sample of AGB stars is not large enough to draw any strong conclusions about the proportion of second generation stars found on the AGB in M53. However, we do observe both first (2) and second (7) generation stars on the AGB, so while the ratio in populations may be different from the RGB, neither population is completely absent.
\item We determine a rate of change for [C/Fe] and [N/Fe] as a function of M$_\mathrm{V}$ for each population to study the evolutionary effects on C and N caused by extra mixing. Each population in M53 has a similar mixing rate for both C and N. Stars in M53 also have a similar depletion rate for C compared to other clusters of similar metallicity, M92 and M15.
\end{enumerate}

\section{Acknowledgements}
We would like to thank Roger A. Bell for making the SSG program available to us. We would also like to thank Zachary Maas and Nicholas Barth for their help in obtaining the stellar spectra during various runs at WIYN. Additionally, we would like to thank Dianne Harmer, Daryl Willmarth, and all of the observing assistants without whom our observations would not have been possible. 

This publication makes use of data products from the Two Micron All Sky Survey, which is a joint project of the University of Massachusetts and the Infrared Processing and Analysis Center/California Institute of Technology, funded by the National Aeronautics and Space Administration and the National Science Foundation. Enrico Vesperini acknowledges support from NSF grant AST-2009193.


\begin{thebibliography}{}
\bibitem[Akaike(1974)]{akaike74} Akaike, H. 1974, ITAC, 19, 716
\bibitem[Alonso et al.(1998)]{tcs} Alonso, A., Arribas, S., \& Martinez-Roger, C.\ 1998, \aaps, 131, 209
\bibitem[Alonso et al.(1999)]{alonso99} Alonso, A., Arribas, S., \& Mart{\'{\i}}nez-Roger, C.\ 1999, \aaps, 140, 261 
\bibitem[Alonso et al.(2001)]{alonso01} Alonso, A., Arribas, S., \& Mart{\'{\i}}nez-Roger, C.\ 2001, \aap, 376, 1039 
\bibitem[Angelou et al.(2012)]{angelou2012} Angelou, G.~C., Stancliffe, R.~J., Church, R.~P., Lattanzio, J.~C., \& Smith, G.~H.\ 2012, \apj, 749, 128 
\bibitem[Bastian \& Lardo(2018)]{bastian2018} Bastian, N. \& Lardo, C.\ 2018, \araa, 56, 83. doi:10.1146/annurev-astro-081817-051839
\bibitem[Baumgardt et al.(2019)]{baumgardt19} Baumgardt, H., Hilker, M., Sollima, A., et al.\ 2019, \mnras, 482, 5138
\bibitem[Bekki(2010)]{bekki2010} Bekki, K.\ 2010, \apjl, 724, L99 
\bibitem[Bell et al.(1994)]{ssg} Bell, R.~A., Paltoglou, G., \& Tripicco, M.~J.\ 1994, \mnras, 268, 771 
\bibitem[Boberg et al.(2016)]{bobergm53} Boberg, O.~M., Friel, E.~D., \& Vesperini, E.\ 2016, \apj, 824, 5 
\bibitem[Briley et al.(2004a)]{briley2004a} Briley, M.~M., Cohen, J.~G., \& Stetson, P.~B.\ 2004a, \aj, 127, 1579 
\bibitem[Briley et al.(2004b)]{briley2004b} Briley, M.~M., Harbeck, D., Smith, G.~H., \& Grebel, E.~K.\ 2004b, \aj, 127, 1588 
\bibitem[Busso et al.(2007)]{busso07} Busso, M., Wasserburg, G.~J., Nollett, K.~M., \& Calandra, A.\ 2007, \apj, 671, 802 
\bibitem[Calura et al.(2019)]{calura19} Calura, F., D'Ercole, A., Vesperini, E., et al.\ 2019, \mnras, 489, 3269
\bibitem[Campbell et al.(2013)]{campbell13} 
Campbell, S.~W., D'Orazi, V., Yong, D., et al.\ 2013, \nat, 498, 198
\bibitem[Campbell et al.(2012)]{campbell12} 
Campbell, S.~W., Yong, D., Wylie-de Boer, E.~C., et al.\ 2012, Galactic Archaeology: Near-Field Cosmology and the Formation of the Milky Way, 458, 205 
\bibitem[Carretta et al.(2009a)]{gir} Carretta, E., Bragaglia, A., Gratton, R.~G., et al.\ 2009a, \aap, 505, 117 
\bibitem[Carretta et al.(2009b)]{uves} Carretta, E., Bragaglia, A., Gratton, R., \& Lucatello, S.\ 2009b, \aap, 505, 139
\bibitem[Chanam{\'e} et al.(2005)]{chaname05} Chanam{\'e}, J., Pinsonneault, M., \& Terndrup, D.~M.\ 2005, \apj, 631, 540
\bibitem[Charbonnel \& Zahn(2007)]{charbonnel07} Charbonnel, C., \& Zahn, J.-P.\ 2007, \aap, 467, L15  
\bibitem[Chun et al.(2010)]{chun2010} Chun, S.-H., Kim, J.-W., Sohn, S.~T., et al.\ 2010, \aj, 139, 606
\bibitem[Chun et al.(2020)]{chun2020} Chun, S.-H., Lee, J.-J., \& Lim, D.\ 2020, arXiv:2008.10410
\bibitem[Cohen(1999a)]{cohen99a} Cohen, J.~G.\ 1999a, \aj, 117, 2428 
\bibitem[Cohen(1999b)]{cohen99b} Cohen, J.~G.\ 1999b, \aj, 117, 2434
\bibitem[C{\^o}t{\'e} et al.(1997)]{cote97} C{\^o}t{\'e}, P., Hanes, D.~A., McLaughlin, D.~E., et al.\ 1997, \apjl, 476, L15
\bibitem[Cowley \& Crampton(1985)]{cowley85} Cowley, A.~P. \& Crampton, D.\ 1985, \pasp, 97, 835
\bibitem[Da Costa(2016)]{dacosta16} Da Costa, G.~S.\ 2016, The General Assembly of Galaxy Halos: Structure, Origin and Evolution, 317, 110
\bibitem[D'Antona et al.(2016)]{dantona2016} D'Antona, F., Vesperini, E., D'Ercole, A., et al.\ 2016, \mnras, 458, 2122
\bibitem[Decressin et al.(2007a)]{decressin2007a} Decressin, T., Charbonnel, C., \& Meynet, G.\ 2007a, \aap, 475, 859 
\bibitem[Decressin et al.(2007b)]{decressin2007b} Decressin, T., Meynet, G., Charbonnel, C., Prantzos, N., \& Ekstr{\"o}m, S.\ 2007b, \aap, 464, 1029 
\bibitem[de Mink et al.(2009)]{demink2009} de Mink, S.~E., Pols, O.~R., Langer, N., \& Izzard, R.~G.\ 2009, \aap, 507, L1 
\bibitem[Denissenkov \& Hartwick(2014)]{denissenkov2014} Denissenkov, P.~A., \& Hartwick, F.~D.~A.\ 2014, \mnras, 437, L21 
\bibitem[Denissenkov \& Tout(2000)]{denissenkov00} Denissenkov, P.~A., \& Tout, C.~A.\ 2000, \mnras, 316, 395 
\bibitem[D'Ercole et al.(2012)]{dercole12} D'Ercole, A., D'Antona, F., Carini, R., et al.\ 2012, \mnras, 423, 1521
\bibitem[D'Ercole et al.(2008)]{dercole2008} D'Ercole, A., Vesperini, E., D'Antona, F., McMillan, S.~L.~W., \& Recchi, S.\ 2008, \mnras, 391, 825 
\bibitem[Dickens(1972)]{dickens72} Dickens, R.~J.\ 1972, \mnras, 159, 7P
\bibitem[Ebrahimi et al.(2020)]{ebrahimi20} Ebrahimi, H., Sollima, A., Haghi, H., et al.\ 2020, \mnras, 494, 4226
\bibitem[Eggleton et al.(2006)]{eggleton2006} Eggleton, P.~P., Dearborn, D.~S.~P., \& Lattanzio, J.~C.\ 2006, Science, 314, 1580 
\bibitem[Eggleton et al.(2008)]{eggleton2008} Eggleton, P.~P., Dearborn, D.~S.~P., \& Lattanzio, J.~C.\ 2008, \apj, 677, 581-592 
\bibitem[Elmegreen(2017)]{elmegreen17} Elmegreen, B.~G.\ 2017, \apj, 836, 80
\bibitem[Forbes \& Bridges(2010)]{forbes10} Forbes, D.~A. \& Bridges, T.\ 2010, \mnras, 404, 1203
\bibitem[Garc{\'{\i}}a-Hern{\'a}ndez et al.(2015)]{garcia15} 
Garc{\'{\i}}a-Hern{\'a}ndez, D.~A., M{\'e}sz{\'a}ros, S., Monelli, M., et al.\ 2015, \apjl, 815, L4
\bibitem[Gerber et al.(2019)]{gerber19} 
Gerber, J.~M., Briley, M.~M., \& Smith, G.~H.\ 2019, \aj, 157, 154
\bibitem[Gerber et al.(2020)]{gerber20} Gerber, J.~M., Friel, E.~D., \& Vesperini, E.\ 2020, \aj, 159, 50 (G20)
\bibitem[Gerber et al.(2018)]{gerber18} Gerber, J.~M., Friel, E.~D., \& Vesperini, E.\ 2018, \aj, 156, 6 (G18)
\bibitem[Gieles et al.(2018)]{gieles18} Gieles, M., Charbonnel, C., Krause, M.~G.~H., et al.\ 2018, \mnras, 478, 2461
\bibitem[Gratton et al.(2019)]{gratton19} Gratton, R., Bragaglia, A., Carretta, E., et al.\ 2019, \aapr, 27, 8
\bibitem[Gratton et al.(2012)]{gratton12} Gratton, R.~G., Carretta, E., \& Bragaglia, A.\ 2012, \aapr, 20, 50 
\bibitem[Gustafsson et al.(2008)]{gustafsson08} Gustafsson, B., Edvardsson, B., Eriksson, K., et al.\ 2008, \aap, 486, 951 
\bibitem[Harbeck et al.(2003)]{bands} Harbeck, D., Smith, G.~H., \& Grebel, E.~K.\ 2003, \aj, 125, 197
\bibitem[Harding(1962)]{harding62} Harding, G.~A.\ 1962, The Observatory, 82, 205
\bibitem[Harris(1996)]{harris} Harris, W.~E.\ 1996, \aj, 112, 1487 
\bibitem[Henkel et al.(2017)]{henkel17} 
Henkel, K., Karakas, A.~I., \& Lattanzio, J.~C.\ 2017, \mnras, 469, 4600
\bibitem[Hubbard \& Dearborn(1980)]{hubbard80} Hubbard, E.~N., \& Dearborn, D.~S.~P.\ 1980, \apj, 239, 248
\bibitem[Johnson et al.(2005)]{johnson05} Johnson, C.~I., Kraft, R.~P., Pilachowski, C.~A., et al.\ 2005, \pasp, 117, 1308 
\bibitem[Johnson et al.(2015a)]{johnson15} 
Johnson, C.~I., McDonald, I., Pilachowski, C.~A., et al.\ 2015a, \aj, 149, 71 
\bibitem[Johnson et al.(2015b)]{johnson15b} Johnson, C.~I., Rich, R.~M., Pilachowski, C.~A., et al.\ 2015b, \aj, 150, 63
\bibitem[Jordi \& Grebel(2010)]{jordi10} Jordi, K. \& Grebel, E.~K.\ 2010, \aap, 522, A71
\bibitem[Keenan(1993)]{keenan93} Keenan, P.~C.\ 1993, \pasp, 105, 905
\bibitem[Kimmig et al.(2015)]{kimmig15} Kimmig, B., Seth, A., Ivans, I.~I., et al.\ 2015, \aj, 149, 53 
\bibitem[Kippenhahn et al.(1980)]{kippenhahn80} 
Kippenhahn, R., Ruschenplatt, G., \& Thomas, H.-C.\ 1980, \aap, 91, 175
\bibitem[Kirby et al.(2015)]{kirby15} Kirby, E.~N., Guo, M., Zhang, A.~J., et al.\ 2015, \apj, 801, 125
\bibitem[Kraft(1994)]{kraft94} Kraft, R.~P.\ 1994, \pasp, 106, 553 
\bibitem[Lagioia et al.(2021)]{lagioia2021} Lagioia, E.~P., Milone, A.~P., Marino, A.~F., et al.\ 2021, arXiv:2101.09843
\bibitem[Lardo et al.(2012)]{lardo12} Lardo, C., Pancino, E., Mucciarelli, A., et al.\ 2012, \aap, 548, A107
\bibitem[Lauchner et al.(2006)]{lauchner06} Lauchner, A., Powell, W.~L., \& Wilhelm, R.\ 2006, \apjl, 651, L33
\bibitem[MacLean et al.(2016)]{maclean16} MacLean, B.~T., Campbell, S.~W., De Silva, G.~M., et al.\ 2016, \mnras, 460, L69 
\bibitem[MacLean et al.(2018)]{maclean18} MacLean, B.~T., Campbell, S.~W., De Silva, G.~M., et al.\ 2018, \mnras, 475, 257 
\bibitem[Marino et al.(2015)]{marino15} Marino, A.~F., Milone, A.~P., Karakas, A.~I., et al.\ 2015, \mnras, 450, 815
\bibitem[Martell et al.(2008a)]{martell} Martell, S.~L., Smith, G.~H., \& Briley, M.~M.\ 2008a, \pasp, 120, 839 
\bibitem[Martell et al.(2008b)]{martellmixing} Martell, S.~L., Smith, G.~H., \& Briley, M.~M.\ 2008b, \aj, 136, 2522
\bibitem[McClure \& Norris(1977)]{mcclure77} McClure, R.~D. \& Norris, J.\ 1977, \apjl, 217, L101
\bibitem[McClure \& Woodsworth(1990)]{mcclure90} McClure, R.~D. \& Woodsworth, A.~W.\ 1990, \apj, 352, 709
\bibitem[McLaughlin \& van der Marel(2005)]{mclaughlin05} McLaughlin, D.~E. \& van der Marel, R.~P.\ 2005, \apjs, 161, 304
\bibitem[Milone et al.(2017)]{miloneatlas} Milone, A.~P., Piotto, G., Renzini, A., et al.\ 2017, \mnras, 464, 3636 
\bibitem[Naidu et al.(2020)]{naidu20} Naidu, R.~P., Conroy, C., Bonaca, A., et al.\ 2020, arXiv:2006.08625
\bibitem[Nardiello et al.(2018)]{nardiello2018} Nardiello, D., Libralato, M., Piotto, G., et al.\ 2018, Monthly Notices of the Royal Astronomical Society, 481, 3382
\bibitem[Nataf et al.(2013)]{lfb} Nataf, D.~M., Gould, A.~P., Pinsonneault, M.~H., \& Udalski, A.\ 2013, \apj, 766, 77
\bibitem[Nataf et al.(2019)]{nataf19} Nataf, D.~M., Wyse, R.~F.~G., Schiavon, R.~P., et al.\ 2019, \aj, 158, 14
\bibitem[Nordhaus et al.(2008)]{nordhaus08} Nordhaus, J., Busso, M., Wasserburg, G.~J., Blackman, E.~G., \& Palmerini, S.\ 2008, \apjl, 684, L29 
\bibitem[Norris et al.(1981)]{norris81} Norris, J., Cottrell, P.~L., Freeman, K.~C., \& Da Costa, G.~S.\ 1981, \apj, 244, 205 
\bibitem[Norris \& Freeman(1979)]{norris79} Norris, J. \& Freeman, K.~C.\ 1979, \apjl, 230, L179
\bibitem[Palacios et al.(2006)]{palacios06} Palacios, A., Charbonnel, C., Talon, S., \& Siess, L.\ 2006, \aap, 453, 261 
\bibitem[Palmerini et al.(2009)]{palmerini09} Palmerini, S., Busso, M., Maiorca, E., \& Guandalini, R.\ 2009, \pasa, 26, 161 
\bibitem[Pancino et al.(2010)]{pancino2010} Pancino, E., Rejkuba, M., Zoccali, M., et al.\ 2010, \aap, 524, A44. doi:10.1051/0004-6361/201014383
\bibitem[Pavlenko et al.(2003)]{pavlenko} Pavlenko, Y.~V., Jones, H.~R.~A., \& Longmore, A.~J.\ 2003, \mnras, 345, 311 
\bibitem[Piotto et al.(2015)]{piotto} Piotto, G., Milone, A.~P., Bedin, L.~R., et al.\ 2015, \aj, 149, 91
\bibitem[Prantzos \& Charbonnel(2006)]{prantzos2006} Prantzos, N., \& Charbonnel, C.\ 2006, \aap, 458, 135 
\bibitem[Rey et al.(1998)]{rey98} Rey, S.-C., Lee, Y.-W., Byun, Y.-I., et al.\ 1998, \aj, 116, 1775
\bibitem[Schwarz(1978)]{schwarz78} Schwarz, G. 1978, AnSta, 6, 461
\bibitem[Sharina et al.(2012)]{sharina12} Sharina, M., Aringer, B., Davoust, E., et al.\ 2012, \mnras, 426, L31
\bibitem[Skrutskie et al.(2006)]{2mass} Skrutskie, M.~F., Cutri, R.~M., Stiening, R., et al.\ 2006, \aj, 131, 1163
\bibitem[Smith \& Martell(2003)]{m3cfe} Smith, G.~H., \& Martell, S.~L.\ 2003, \pasp, 115, 1211 
\bibitem[Smith \& Mateo(1990)]{smith90} Smith, G.~H. \& Mateo, M.\ 1990, \apj, 353, 533
\bibitem[Smith et al.(2013)]{smith2013} Smith, G.~H., Modi, P.~N., \& Hamren, K.\ 2013, \pasp, 125, 1287 
\bibitem[Smith \& Norris(1982)]{smith82} Smith, G.~H. \& Norris, J.\ 1982, \apj, 254, 149
\bibitem[Smith(2015)]{smith2015} Smith, G.~H.\ 2015, \pasp, 127, 825 
\bibitem[Smolinski et al.(2011)]{smolinski11} Smolinski, J.~P., Martell, S.~L., Beers, T.~C., et al.\ 2011, \aj, 142, 126
\bibitem[Suntzeff \& Smith(1991)]{suntzeff91} Suntzeff, N.~B., \& Smith, V.~V.\ 1991, \apj, 381, 160 
\bibitem[Sweigart \& Mengel(1979)]{sweigart79} Sweigart, A.~V., \& Mengel, J.~G.\ 1979, \apj, 229, 624 
\bibitem[Trefzger et al.(1983)]{trefzger83} 
Trefzger, D.~V., Langer, G.~E., Carbon, D.~F., et al.\ 1983, \apj, 266, 144
\bibitem[Vasiliev(2019)]{vasiliev2019} Vasiliev, E.\ 2019, \mnras, 484, 2832
\bibitem[Ventura et al.(2001)]{ventura2001} Ventura, P., D'Antona, F., Mazzitelli, I., \& Gratton, R.\ 2001, \apjl, 550, L65 
\bibitem[Vesperini et al.(2013)]{vesperini2013} Vesperini, E., McMillan, S.~L.~W., D'Antona, F., \& D'Ercole, A.\ 2013, \mnras, 429, 1913 
\bibitem[Yuan et al.(2020)]{yuan20} Yuan, Z., Chang, J., Beers, T.~C., et al.\ 2020, \apjl, 898, L37
\bibitem[Yong et al.(2015)]{yong15} Yong, D., Grundahl, F., \& Norris, J.~E.\ 2015, \mnras, 446, 3319
\end{thebibliography}
\end{document}